\documentclass{ecai}
\usepackage{graphicx}
\usepackage{latexsym}
\usepackage{graphicx}
\usepackage{amsmath}
\usepackage{amssymb}
\usepackage{booktabs}
\usepackage{xcolor}
\usepackage{tabularx}
\usepackage{url}
\usepackage{color}

\paperid{1024}

\begin{document}

\begin{frontmatter}

\title{High Dynamic Range Image Reconstruction via Deep Explicit Polynomial Curve Estimation}
\author[A,B,C,D,E]{\fnms{Jiaqi}~\snm{Tang}}
\author[F,G]{\fnms{Xiaogang}~\snm{Xu}}
\author[D]{\fnms{Sixing}~\snm{Hu}}
\author[A,B,C]{\fnms{Ying-Cong}~\snm{Chen} \thanks{Corresponding Author. Email: yingcongchen@ust.hk}}

\address[A]{Hong Kong University of Science and Technology (Guangzhou)}
\address[B]{Hong Kong University of Science and Technology}
\address[C]{HKUST(GZ)-SmartMore Joint Lab}
\address[D]{SmartMore Technology}
\address[E]{Northwestern Polytechnical University}
\address[F]{Zhejiang Lab}
\address[G]{Zhejiang University}

\begin{abstract}
Due to limited camera capacities, digital images usually have a narrower dynamic illumination range than real-world scene radiance. To resolve this problem, High Dynamic Range (HDR) reconstruction is proposed to recover the dynamic range to better represent real-world scenes. However, due to different physical imaging parameters, the tone-mapping functions between images and real radiance are highly diverse, which makes HDR reconstruction extremely challenging. Existing solutions can not explicitly clarify a corresponding relationship between the tone-mapping function and the generated HDR image, but this relationship is vital when guiding the reconstruction of HDR images. To address this problem, we propose a method to explicitly estimate the tone mapping function and its corresponding HDR image in one network. Firstly, based on the characteristics of the tone mapping function, we construct a model by a polynomial to describe the trend of the tone curve. To fit this curve, we use a learnable network to estimate the coefficients of the polynomial. This curve will be automatically adjusted according to the tone space of the Low Dynamic Range (LDR) image, and reconstruct the real HDR image. Besides, since all current datasets do not provide the corresponding relationship between the tone mapping function and the LDR image, we construct a new dataset with both synthetic and real images. Extensive experiments show that our method generalizes well under different tone-mapping functions and achieves SOTA performance. The code/dataset is available at \textcolor{blue}{\textbf{\texttt{https://github.com/jqtangust/EPCE-HDR.git}}}.
\end{abstract}

\end{frontmatter}

%

\section{Introduction}
\label{sec:intro}

\begin{figure*}[t]
  \centering
  \vspace{-0.2in}
  \includegraphics[width=1\textwidth]{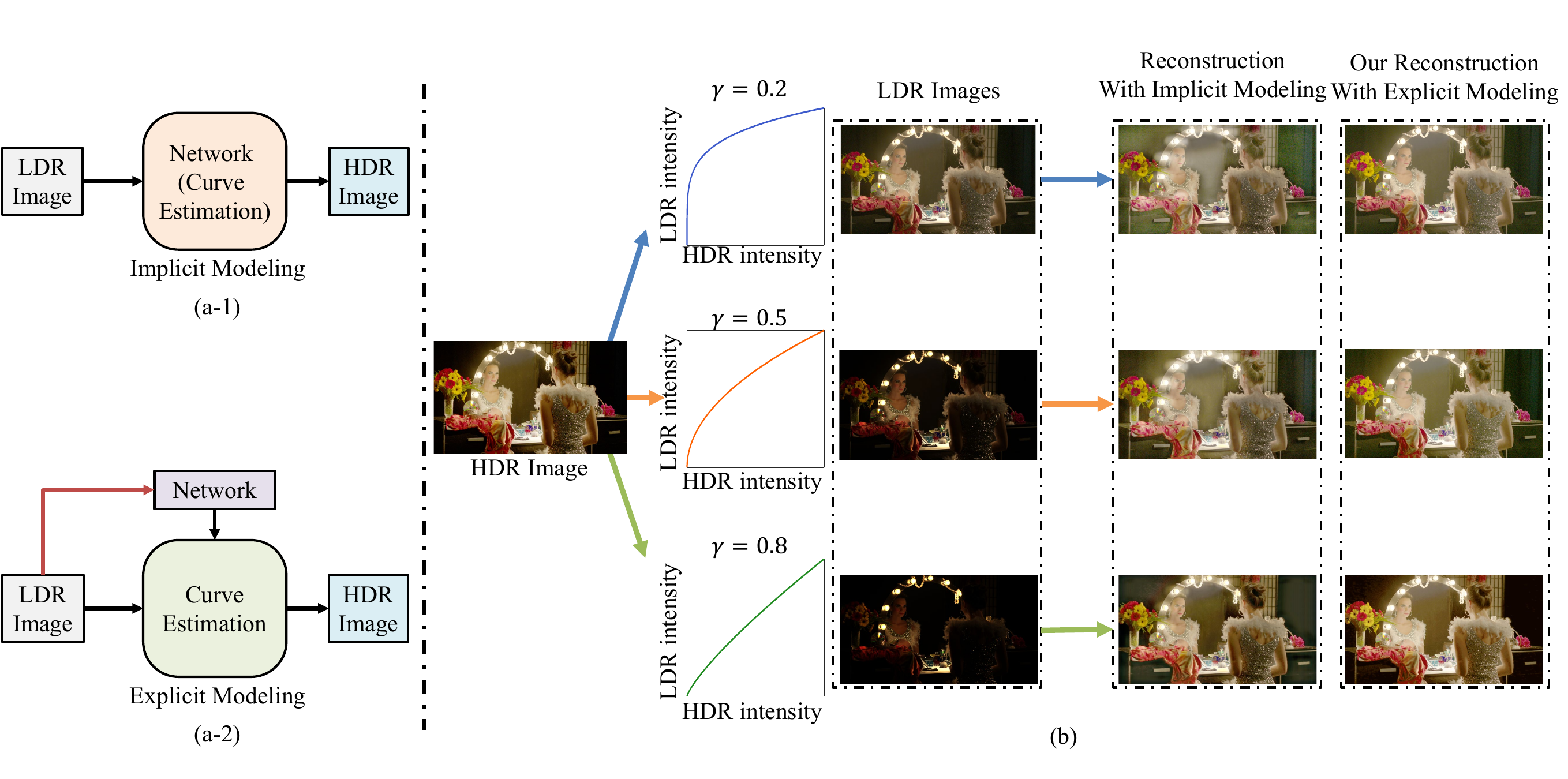}
  \vspace{-0.3in}
  \caption{Traditional HDR reconstruction approaches (a-1) build the mapping relation between LDR and HDR images directly. Without explicit modeling for the corresponding physical process, they can not capture enough information about the tone-mapping function and thus show limited performance in generalizing to HDR reconstruction. We propose a framework to formulate such explicit modeling (a-2) by estimating the tone-mapping curve from the LDR image. As shown in (b), three vastly varying tone-mapping curves (with the form of $\gamma$-curves) are applied to the HDR image to obtain LDR images. Our framework can satisfactorily restore the HDR image with different tone-mapping degradations, while methods with implicit modeling show different reconstruction effects for diverse curves.}
  \label{fig:d}
\end{figure*}

Real-world scenes largely exceed the working range of consumer-grade camera sensors~\cite{radonjic2011dynamic}. Therefore, cameras have to clip and compress the dynamic ranges, leading to Low Dynamic Range (LDR) of digital images. High Dynamic Range (HDR) reconstruction aims to improve the dynamic range of images for better representing real-world scenes. It has attracted a lot of interest in recent years and has been widely used in television, photography, film production, and industry~\cite{kang2003high,mann2012realtime,mccann2011art,tocci2011versatile,myszkowski2008high}.

A common setting~\cite{eilertsen2017hdr,marnerides2018expandnet,chen2021hdrunet} for HDR reconstruction is to reverse the tone mapping function~\cite{banterle2006inverse}, and these methods~\cite{liu2020single} assume that the tone mapping function used in the LDR image is included in a predefined repository known as the Empirical Model of Response (EMoR)~\cite{grossberg2003space}. Nevertheless, this repository comprises a finite number of tone mapping curves, which do not account for all possible scenarios of the tone mapping function. Consequently, a model that performs well under such conditions~\cite{PerezPellitero2021NTIRE2C} may fail to generalize when confronted with real-world images.

Only a few existing approaches take this into account, Pan et al.~\cite{pan2021metahdr} propose a meta-learning approach that learns to capture the common patterns between different tone-mappings, yet a clear corresponding relationship between LDR image and tone mapping functions is still ambiguous. Most existing learning-based approaches~\cite{eilertsen2017hdr,chen2021hdrunet,marnerides2018expandnet,lee2018deep} capture the information of tone-mapping functions implicitly. However, this does not work well when the tone space changes dramatically, as shown in Figure~\ref{fig:d}.

In this paper, we present a novel framework for HDR reconstruction that can handle a wide range of tone spaces and explicitly estimate the corresponding tone mapping function for each image. e. Firstly, we model the camera pipeline as a nonlinear tone-mapping followed by a degeneration model. The goal of HDR reconstruction is to reverse this process by recovering degenerated information and inverting the nonlinear mapping. Our focus in this paper is on the latter part, i.e., how to train a model to reverse an unknown nonlinear tone-mapping.

To address this problem, we propose to parameterize the unknown tone-mapping function using the polynomial function family and estimate its coefficients with a Pyramid-Path Vision Transformer (PPViT). PPViT is an architecture that can effectively leverage multi-scale information from LDR images. This is particularly important for inverse tone-mapping function estimation, as multi-scale information is critical for accurately reconstructing the HDR image~\cite{lin2004radiometric,matsushita2007radiometric,li2017radiometric}.

To assess the efficacy of our approach, we have developed a novel dataset that can simulate real-world scenarios with a wide range of tone-mapping functions. Our experimental results demonstrate that our model performs exceptionally well on a diverse set of images and effectively captures the corresponding relationships across various tone-mapping functions.

Our contributions are summarized as follows:
\vspace{-0.1in}
\begin{itemize}


\item We propose a new framework for explicitly modeling and estimating tone-mapping function parameters, which is capable of handling diverse tone-mapping functions, providing more accurate reconstructions in the process.
\item We have constructed a new dataset specifically designed for this task, featuring a clear relationship between LDR images and their corresponding tone mapping functions.
\item Our approach achieves SOTA performance in both synthesis and real dataset and clarifies a corresponding relationship between the tone-mapping function and the generated HDR image.
\end{itemize}

\section{Related Works}
\label{sec:related}

\noindent \textbf{High Dynamic Range Reconstruction} In general, HDR reconstruction can be divided into two categories: single-frame HDR image reconstruction (Si-HDR), and multi-frame HDR reconstruction (MF-HDR).
MF-HDR focuses on the different information of images with different exposures~\cite{yan2019attention}, and blends this information to obtain high dynamic range images.
In contrast, Si-HDR aims to recover information of dynamic range from a single image, which is even more challenging.
This paper mainly focuses on Si-HDR. Eilertsen et al.~\cite{eilertsen2017hdr} first introduced deep learning into HDR image reconstruction and solved the problem of predicting information that has been lost from a single exposure.
Endo et al.~\cite{Endo2017DeepRT} merged the LDR images with different exposures (i.e., the images in brackets) and then reconstructed the HDR images by merging them.
Marnerides et al.~\cite{marnerides2018expandnet} designed a multi-scale architecture to improve image quality.
These methods always assume that training and testing are on the same type of tone curve so they do not generalize well when faced with different tone curves.
Lee et al.~\cite{Lee2018DeepRH} and Raipurkar et al.~\cite{Raipurkar2021HDRcGANSL} were trying to propose a generative approach to solve the HDR reconstruction problem, but the core of their problem is still the recovery of the exposed area.
Chen et al.~\cite{chen2021hdrunet} considered that noise and quantization errors are often generated during the imaging process, so they use a U-shape network to focus on denoising and dequantization.
Liu et al.~\cite{liu2020single} proposed an interesting method to learn a reversed camera pipeline in three different independent stages. In their study, they used a simple method based on PCA and EMoR models to estimate the CRF, but this would limit the space of CRF to EMoR.
In summary, most previous methods do not model the unknown tone-mapping function, and may encounter generalization problems in practical scenarios when training and testing images have very different tone-mapping functions. Our approach is intrinsically different by explicitly estimating the tone-mapping function, which is advantageous for diverse camera parameters.

\noindent \textbf{Blind Image Restoration} Our work can be categorized into blind image restoration. Blind image restoration has been investigated in super-resolution~\cite{Liu2021BlindIS,Wang2021RealESRGANTR,Wang2021UnsupervisedDR}, denoising~\cite{Guo2019TowardCB,Cha2021GAN2GANGN} and deblurring~\cite{Wen2021ASL}.

Blind image super-resolution~\cite{Liu2021BlindIS} generally estimates kernels and noise levels for the degeneration process, allowing the model to deal with real-world images with different types of degeneration. Similarly, blind denoising~\cite{Guo2019TowardCB} generally includes a noise estimation sub-model in the pipeline, which avoids overfitting specific types of noise. Our work can be viewed as a blind problem in the sense that we explicitly estimate the tone-mapping function. This allows our model to adapt to LDR images processed with different tone mapping functions.

\section{Methods}

\begin{figure*}
  \centering
  \includegraphics[width=1\textwidth]{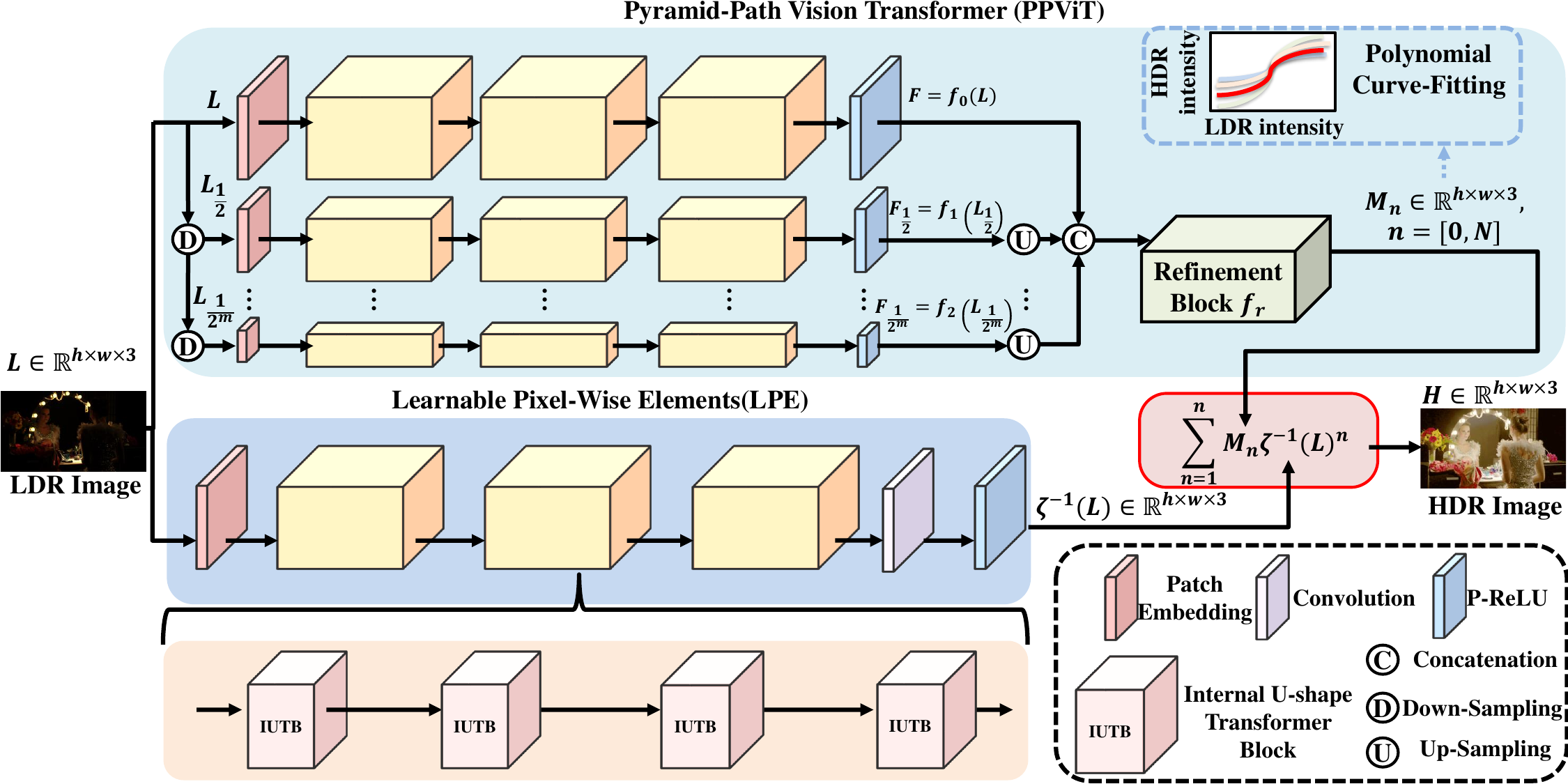}
  \vspace{-0.2in}
  \caption{The overall architecture of our framework, where we estimate the polynomial coefficient maps from the input LDR image that represent the tone-mapping curves.}
  \vspace{-0.1in}
  \label{fig:overall}
\end{figure*}

In general, a digital camera captures images by firstly recording scene radiance to HDR images, then converting HDR images to LDR ones. LDR images are displayable on most screens, and are extremely rich in our daily life. We aim to invert LDR images back to HDR ones, so that photo retouching and enhancement are more achievable. 

The HDR-to-LDR conversion generally includes dynamic range clipping, tone mapping, and quantization \cite{liu2020single}. 
Dynamic range clipping clips HDR values to a limited range. Tone mapping adjusts pixel values to the displayable range while preserving image contents that are important to the scene. 
Quantization discretizes real pixel values to discretized ones (usually 8 bits). 
Here, dynamic range clipping and quantization are degeneration processes that inevitably cause information loss, while tone mapping is a non-linear function that could vary for different cameras or different artists. With this, the HDR-to-LDR conversion can be simplified as 
\begin{equation} \label{eq:HDR2LDR}
    L=\zeta(\tau (H)),
\end{equation}
where $H \in \mathbb{R}^{h\times w \times c}$ and $L \in \mathbb{R}^{h\times w \times c}$ are HDR and LDR images respectively, $\tau(\cdot)$ is the tone-mapping function, and $\zeta(\cdot)$ models the degradation process caused by dynamic range clipping and quantization. The inverse conversion, i.e., HDR reconstruction, can thus be formulated as
\begin{equation} \label{eq:LDR2HDR}
    H=\tau^{-1}(\zeta^{-1} (L)).
\end{equation}
Compared with existing works \cite{chen2021hdrunet,marnerides2018expandnet} that treat the conversion between HDR and LDR as a whole, this formulation decomposes degeneration and tone-mapping, allowing for modeling these two processes separately. 

Our overall framework is illustrated in Fig.~\ref{fig:overall}. We model $\tau^{-1}(\cdot)$ and $\zeta^{-1}(\cdot)$ with two separate modules, i.e., the Pyramid-Path Vision Transformer (PPViT) and Learnable Pixel-wise Elements (LVE). These two modules are then combined to generate the final output by Deep Polynomial Curve Estimation. In our implementation, $\zeta^{-1}(\cdot)$ is basically a transformer whose implementation details are shown in the supplementary file. $\tau^{-1}(\cdot)$ is carefully designed, allowing for handling unknown tone-mappings. The key idea will be elaborated as follows. 

\subsection{Parameterizing the Tone-Mapping Function}

Although the inversed tone-mapping functions $\tau^{-1}(\cdot)$ could vary for different images, they generally follow some common structures, i.e., they are usually monotonic or at least semi-monotonic~\cite{grossberg2004modeling}. In this sense, we can use certain function families to parameterize tone-mapping functions. In this work, we choose the polynomial function following~\cite{mitsunaga1999radiometric}, and thus Eq. \eqref{eq:LDR2HDR} can be rewritten as
\begin{equation}
    H =\tau^{-1}(\zeta^{-1} (L)) = \sum_{n=0}^N {M_n \zeta^{-1}(L)^n},
    \label{eq:poly}
\end{equation}
where $M_n \in \mathbb{R}^{h\times w}, n\in[0, N]$ are Polynomial Coefficient Maps (PCMs), $N$ is the degree of polynomial, and $\zeta^{-1}(L)^n$ means taking power $n$ for all pixels of $\zeta^{-1}(L)^n$. Therefore, learning $\tau^{-1}(\cdot)$ is equivalent to learning the coefficients $M_n$.

\subsection{Learning the Polynomial Coefficient Maps}

Even after representing $\tau^{-1}(\cdot)$ as PCMs, learning $\tau^{-1}(\cdot)$ is still challenging. During inference time, the model needs to infer tone-mapping functions based purely on LDR images without accessing their HDR counterparts. 
Also, in this paper, we consider the scenario where the tone-mapping functions of testing images are different from training ones. Thus memorizing the training $\tau(\cdot)$ with a neural network may not generalize well. Fortunately, existing works show that the tone-mapping function can be estimated from a single LDR image based on some statistical information. For example, Lin et al.~\cite{lin2004radiometric} show that color distribution at specific edge regions reveals characteristics of the tone-mapping function. Matsushita and Lin \cite{matsushita2007radiometric} prove that tone-mapping can be estimated by the symmetry of noise distribution. Li we al.~\cite{li2017radiometric} demonstrate that human faces contain rich cues for tone-mapping estimation. Although these works are limited in some restricted scenarios, they demonstrate the feasibility of estimating $\tau(\cdot)$ (or equivalently, $\tau^{-1}(\cdot)$) from single LDR images, which motivates our solution.

As traditional approaches \cite{lin2004radiometric,matsushita2007radiometric,li2017radiometric} leverage both local and global information to estimate the tone-mapping function, our model is also designed to capture multi-scale information. As shown in Fig. \ref{fig:overall}, we propose the Pyramid-Path Vision Transformer (PPViT), a new transformer with pyramid paths that fuse image features from different scales. Besides, we also propose the Internal U-shape Transformer Block (IUTB) that enables the transformer to effectively extract local features.

\subsubsection{Pyramid-Path Vision Transformer (PPViT)} 
PPViT is used to extract image features of different scales.
Let $D^m(·)$ denotes the down-sampling operation that transfers $h\times w \times 3$ into $\frac{h}{2^m}\times \frac{w}{2^m} \times 3$. We obtain the pyramid representation of $L$ as

\begin{equation}
\{ L, L_{\frac{1}{2}},...,L_{\frac{1}{2^m}} \},\quad {\rm where}\quad L_{\frac{1}{2^m}}=D^m(L),
\end{equation}
where $\{ L_{\frac{1}{2}},...,L_{\frac{1}{2^m}} \}$ mean the multi-scale images besides the input image $L$. 

Then, the multi-scale information will be processed via our transformer-based paths to extract the corresponding feature, as
\begin{equation}
F=f_0 (L), F_{\frac{1}{2}}=f_1(L_{\frac{1}{2}}), ..., F_{\frac{1}{2^m}}=f_m(L_{\frac{1}{2^m}}),
\end{equation}

where $f_0,...,f_m$ denote different transformer-based paths, and $\{F, F_{\frac{1}{2}},...,F_{\frac{1}{2^m}} \}$ are the extracted features whose shapes are the same shape as $\{L, L_{\frac{1}{2}},...,L_{\frac{1}{2^m}} \}$.
These multi-scale features are also merged via a pyramid manner with up-sampling and concatenation operations. The merged feature will be employed to obtain the prediction of PCMs through a refinement block, as
\begin{equation}
\{ M_0,...M_N \}=f_r(F \oplus U^1(F_{\frac{1}{2}}) ... \oplus U^m(F_{\frac{1}{2^m}})),
\end{equation}
where $\{ M_0,...M_N \}$ are the PCMs in Eq.~\ref{eq:poly}, $U^m$ denotes the up-sampling operation that transfers $\frac{h}{2^m}\times \frac{w}{2^m} \times 3$ into $h\times w \times 3$, $\oplus$ means the concatenation operation, $f_r$ is the refinement block which consists of four Internal U-shape Transformer Blocks and one convolution layer. 
The function of the refinement block is to further promote the fusion of multi-scale features.
After some tests, in order to balance the model parameter and performance of our network, we currently set $m=2$.

\subsubsection{Internal U-shape Transformer Block (IUTB)} 

\begin{figure}
  \centering
  \includegraphics[width=0.5\textwidth]{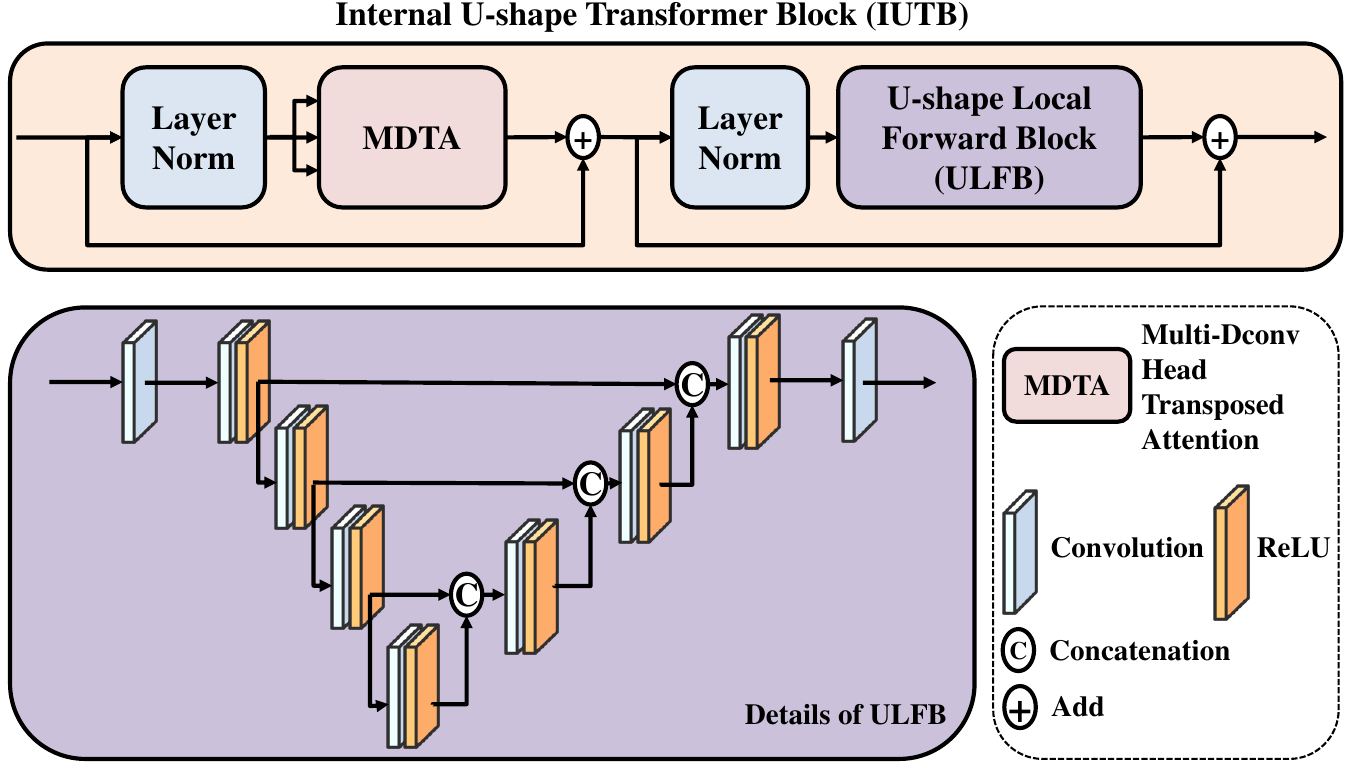}
  \vspace{-0.3in}
  \caption{The illustration of our designed Internal U-shape Transformer Block (IUTB) that enables the transformer to effectively extract local features.}
  \vspace{-0.1in}
  \label{fig:iutb}
\end{figure}

In our PPViT, we replace the commonly used transformer~\cite{vaswani2017attention} with an Internal U-shape Transformer Block (IUTB). Compared with the traditional transformer, IUTB is advantageous in extracting local information which is important in estimating tone-mapping functions \cite{lin2004radiometric}. As shown in Fig.~\ref{fig:iutb}, IUTB is composed of an MDTA module defined in \cite{zamir2021restormer} and a U-shape Local Forward Block that better extracts local features. As shown in Sec. \ref{as}, this design leads to performance improvement. 

\subsection{Details in Implementation}
Our model is trained in an end-to-end manner. Specifically, we use the smooth $L_1$ loss to train our model. Our initial learning rate is set to $10^{-5}$ and the network is iterated in 8 million steps. The learning rate is scheduled in steps, where the learning rate decays to one-half of the original rate after every 2 million steps. The batch size is $4$. 

\begin{figure*}[t]
	\centering
	\Huge
	\newcommand\widthface{0.35}
    \newcommand\widthfaceweight{0.1952631578947369}
	\resizebox{1.0\linewidth}{!}{
	\begin{tabular}{cccccc}
	    LDR & ExpandNet~\cite{marnerides2018expandnet} & HDRUNET~\cite{chen2021hdrunet} & Uformer~\cite{Wang2021UformerAG} & Ours & Ground Truth \\
		\includegraphics[width=\widthface\textwidth,height=\widthfaceweight\textwidth]{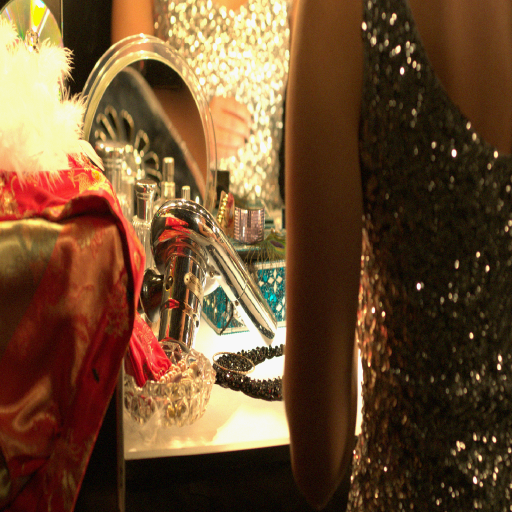} &
		\includegraphics[width=\widthface\textwidth,height=\widthfaceweight\textwidth]{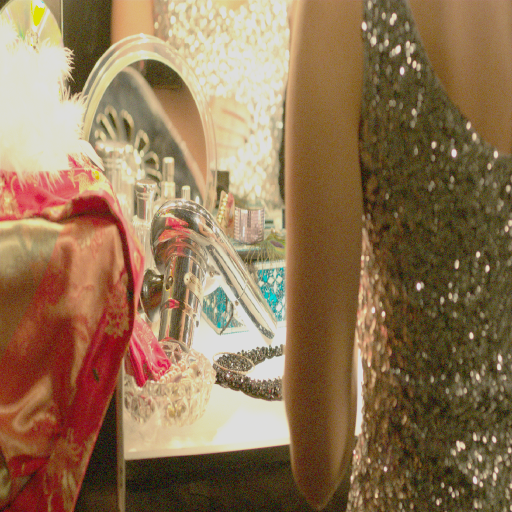} &
		\includegraphics[width=\widthface\textwidth,height=\widthfaceweight\textwidth]{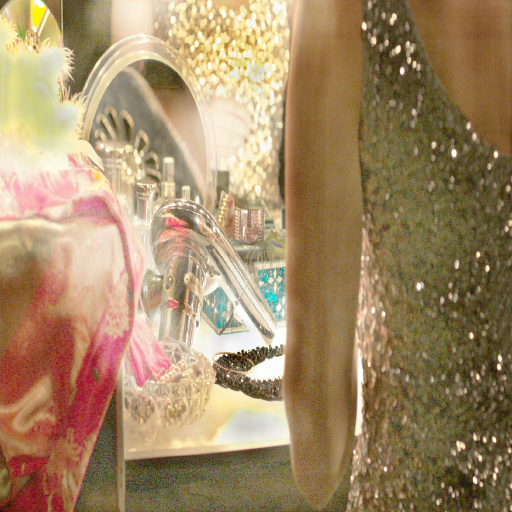} &
		\includegraphics[width=\widthface\textwidth,height=\widthfaceweight\textwidth]{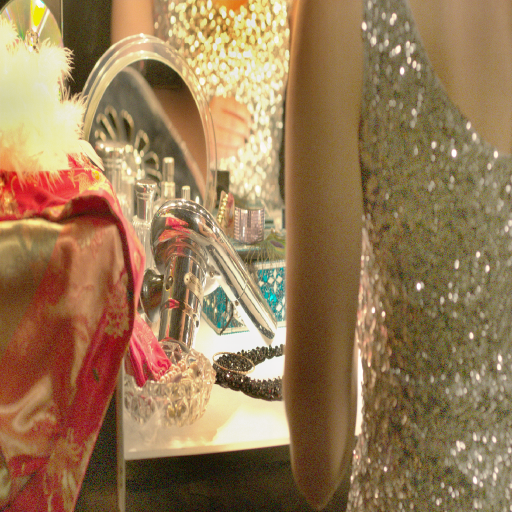} &
		\includegraphics[width=\widthface\textwidth,height=\widthfaceweight\textwidth]{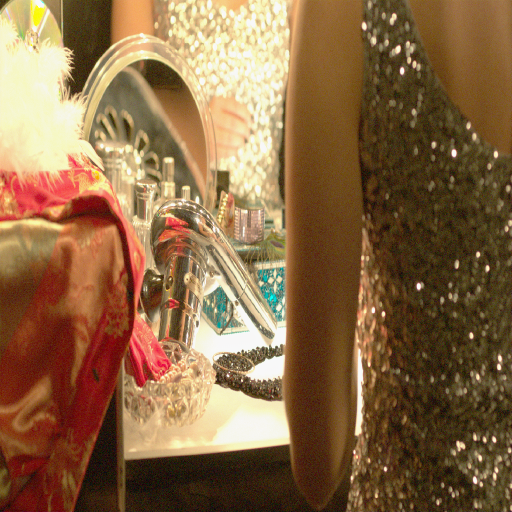} &
		\includegraphics[width=\widthface\textwidth,height=\widthfaceweight\textwidth]{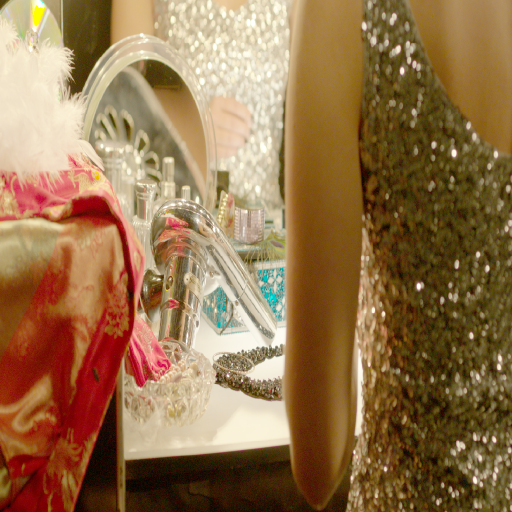} \\
		
		\includegraphics[width=\widthface\textwidth,height=\widthfaceweight\textwidth]{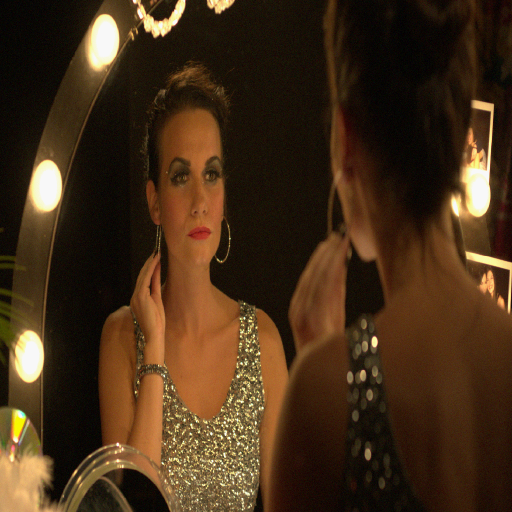} &
		\includegraphics[width=\widthface\textwidth,height=\widthfaceweight\textwidth]{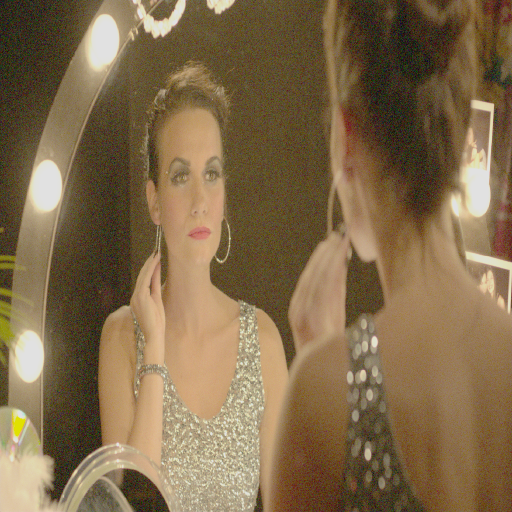} &
		\includegraphics[width=\widthface\textwidth,height=\widthfaceweight\textwidth]{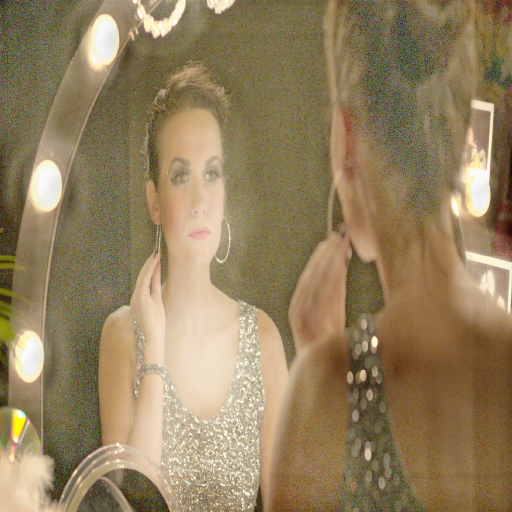} &
		\includegraphics[width=\widthface\textwidth,height=\widthfaceweight\textwidth]{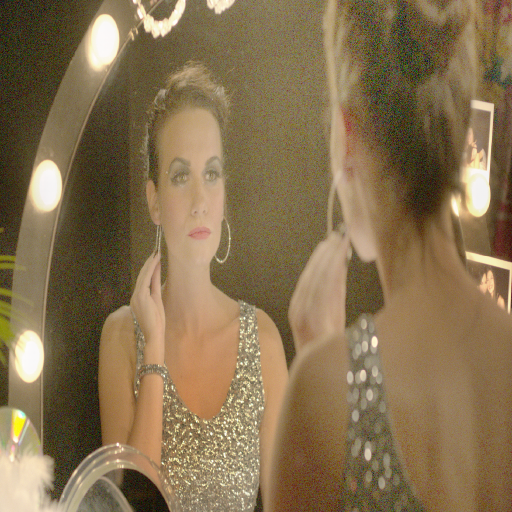} &
		\includegraphics[width=\widthface\textwidth,height=\widthfaceweight\textwidth]{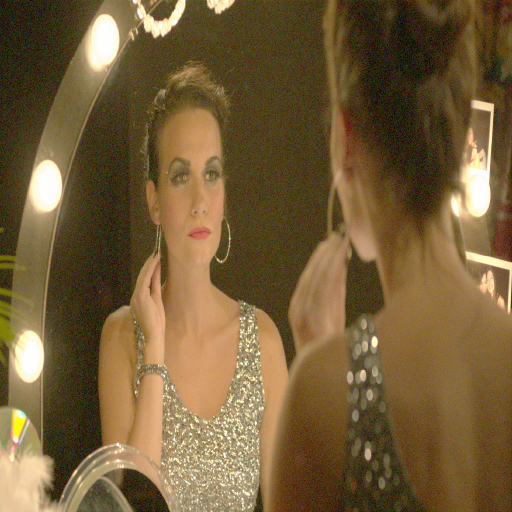} &
		\includegraphics[width=\widthface\textwidth,height=\widthfaceweight\textwidth]{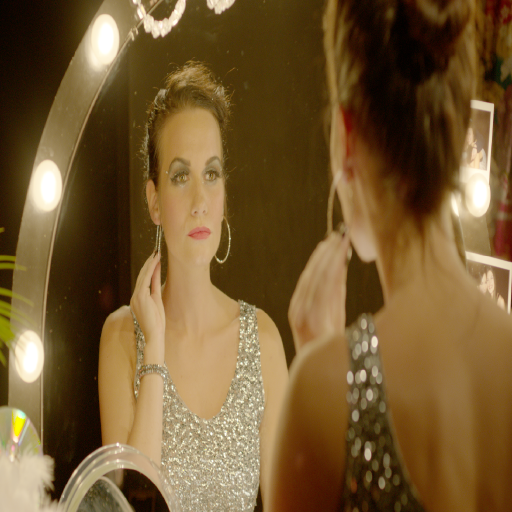} \\
		
		\includegraphics[width=\widthface\textwidth,height=\widthfaceweight\textwidth]{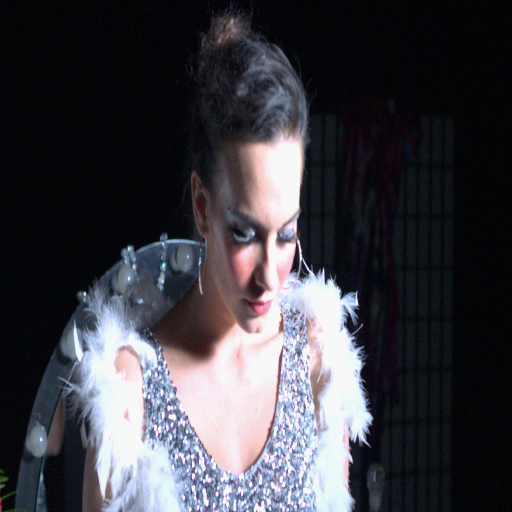} &
		\includegraphics[width=\widthface\textwidth,height=\widthfaceweight\textwidth]{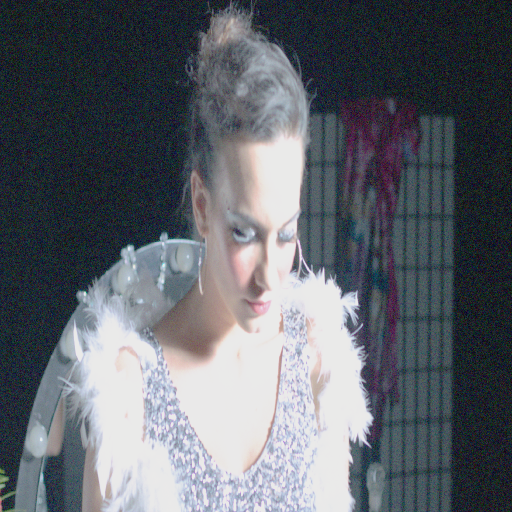} &
		\includegraphics[width=\widthface\textwidth,height=\widthfaceweight\textwidth]{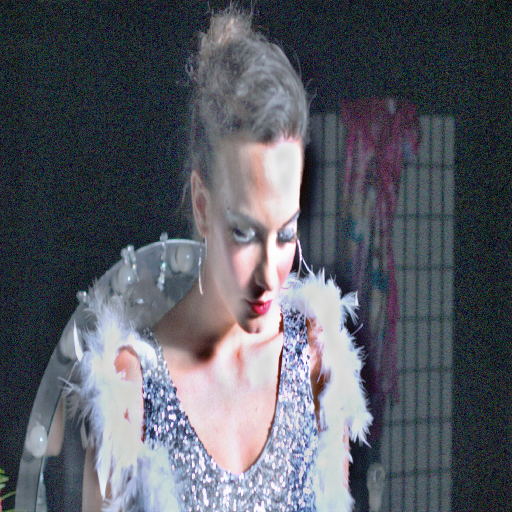} &
		\includegraphics[width=\widthface\textwidth,height=\widthfaceweight\textwidth]{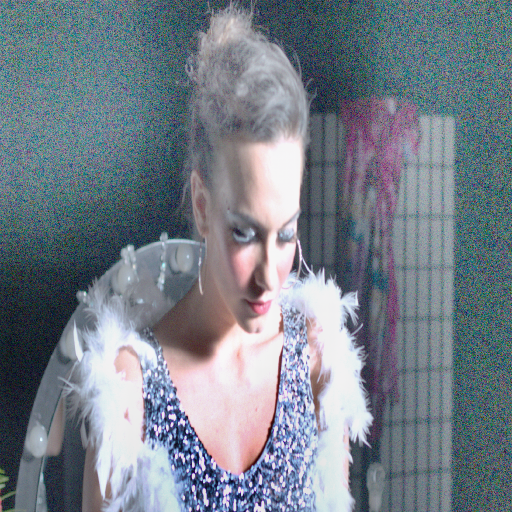} &
		\includegraphics[width=\widthface\textwidth,height=\widthfaceweight\textwidth]{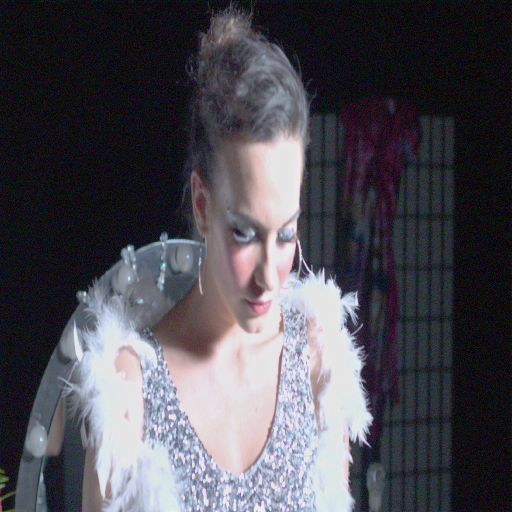} &
		\includegraphics[width=\widthface\textwidth,height=\widthfaceweight\textwidth]{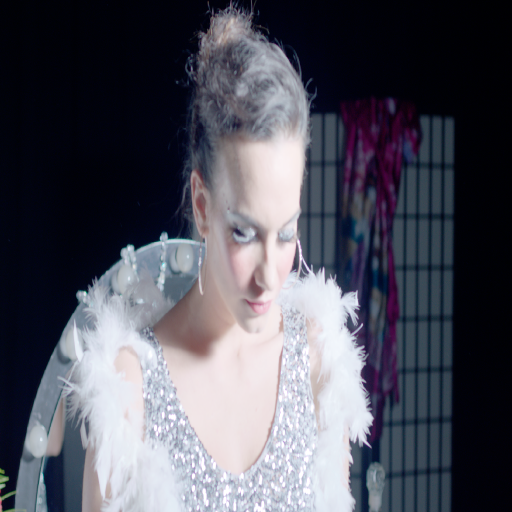} \\
	\end{tabular}}
	\vspace{-0.2in}
	\caption{Visualization comparison in Test-Real dataset. \textcolor[rgb]{ .122,  .306,  .471}{(Photos are in high resolution. Please zoom in to see more details.)}}
 \vspace{-0.2in}
	\label{real}
\end{figure*}

\section{HDR Dataset}

\subsection{Details of the Dataset}
To evaluate the performance of different frameworks on the HDR reconstruction task, we build a new dataset based on the benchmark from the NTIRE HDR Competition~\cite{PerezPellitero2021NTIRE2C}.
As we have stated, the polynomial functions can approximate the real-world tone-mapping functions with the combination of varying polynomial coefficients. Thus, to ensure the trained models can be generalizable to arbitrary LDR images that undergo unknown tone-mapping degradations, we obtain the LDR and HDR image pairs from the original training set~\cite{PerezPellitero2021NTIRE2C} by applying polynomial-based tone-mapping curves on the corresponding HDR images.
In notation, given the HDR image $H$, the LDR images can be acquired as
\begin{equation}
    L=\frac{log\left( 1+\mu H \right)}{log\left( 1+\mu \right)},
\end{equation}
where $\mu$ can be varying from $1$ to $2*10^6$, extending the range of tone-mapping curves. Especially, for one HDR image, we can synthesize 20 LDR images.

On the other hand, the testing set is divided into two parts. The first part is the testing benchmark from the NTIRE HDR Competition~\cite{PerezPellitero2021NTIRE2C}, called ``Test-Real'', which represents real-world degradation. Moreover, to verify whether our method can fit various types of tone-mapping curves, we choose a stack of gamma tone-mapping functions that are largely different from the polynomial functions in the training set, and apply the gamma tone-mapping functions to the HDR images from ~\cite{PerezPellitero2021NTIRE2C}, obtaining the LDR and HDR image pairs.
Given a HDR image $H$, the corresponding LDR image is computed as $L=b\cdot H^{\gamma}$, where $b=1$ and $\gamma \in \left( 0,1 \right]$.
In addition, we set 8 values for $\gamma$ in this testing set that is called ``Test-Gamma''.

\subsection{Training and Testing Strategies}

During training, for each scene, we randomly select one image from the corresponding LDR stack and input it into the network. Then we calculate the smooth $L_1$ loss between the output and the ground truth.
Our test dataset consists of ``Test-Gamma'' and ``Test-real''. For Test-Gamma, we input the synthesized LDR image into our network, validating that our framework can be generalized to other tone-mapping curves. For Test-Real, we recover the HDR images from the real LDR images and calculate quantitative metrics. 

Only 4 Nvidia 3090 GPUs are required to train our model to reach the final version. In the testing phase, we used a single Nvidia 3090 for inference, and the model speed is about 47ms per the HDR reconstruction of the image, which meets the current efficient standard.

\begin{table}[t]
  \centering
  \Huge
  \caption{Quantitative Comparison in Test-Real dataset. \textcolor[rgb]{ 1,  0,  0}{Red} represents the method with the best performance and  \textcolor[rgb]{ .122,  .306,  .471}{{Blue}} represents the second best. }
  \resizebox{1.0\linewidth}{!}{
    \begin{tabular}{cccccccc}
    \toprule
     & PSNR  & $\mu$-PSNR & SSIM & $\mu$-SSIM & HDR-VDP-2.2 & AvgPSNR\\
    \midrule
    ExpandNet~\cite{marnerides2018expandnet} & 10.06 & \textcolor[rgb]{ 1,  0,  0}{22.20} & 0.5203 & \textcolor[rgb]{ .122,  .306,  .471}{0.6498} & 38.48 & 13.70\\
    HDRUNET~\cite{chen2021hdrunet} & \textcolor[rgb]{ .122,  .306,  .471}{27.29} & 17.15 & \textcolor[rgb]{ .122,  .306,  .471}{0.9337} & 0.6199 & 45.73  & \textcolor[rgb]{ .122,  .306,  .471}{24.25} \\
    Uformer~\cite{Wang2021UformerAG} & 26.77 & 16.65 & 0.9109 & 0.6017 & \textcolor[rgb]{ .122,  .306,  .471}{45.99} & 23.74\\
    Ours  & \textcolor[rgb]{ 1,  0,  0}{28.24} & \textcolor[rgb]{ .122,  .306,  .471}{21.49} & \textcolor[rgb]{ 1,  .0,  .0}{0.9406} & \textcolor[rgb]{ 1,  0,  0}{0.6577} & \textcolor[rgb]{ 1,  0,  0}{47.09}  &  \textcolor[rgb]{ 1,  0,  0}{26.23}\\
    \bottomrule
    \end{tabular}}%
  \label{tab:qc}%
\end{table}%

\begin{figure*}[t]
	\centering
	\Huge
	\newcommand\widthface{0.35}
	\newcommand\widthfaceweight{0.1952631578947369}
	\resizebox{1.0\linewidth}{!}{
	\begin{tabular}{cccccccc}
		$\gamma=0.2$ & $\gamma=0.3$ & $\gamma=0.4$ & $\gamma=0.5$& $\gamma=0.6$& $\gamma=0.7$& $\gamma=0.8$& $\gamma=1.0$\\
		\includegraphics[width=\widthface\textwidth,height=\widthfaceweight\textwidth]{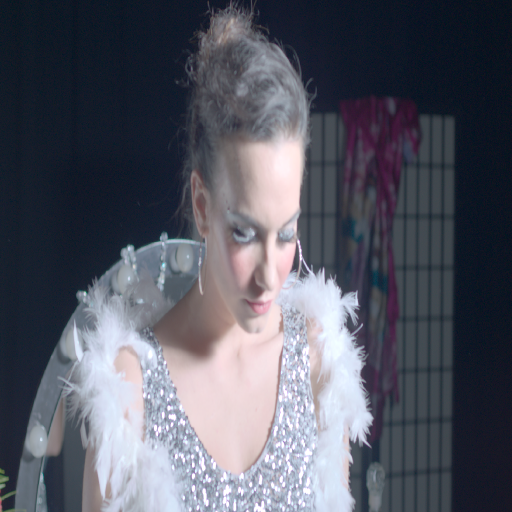} &
		\includegraphics[width=\widthface\textwidth,height=\widthfaceweight\textwidth]{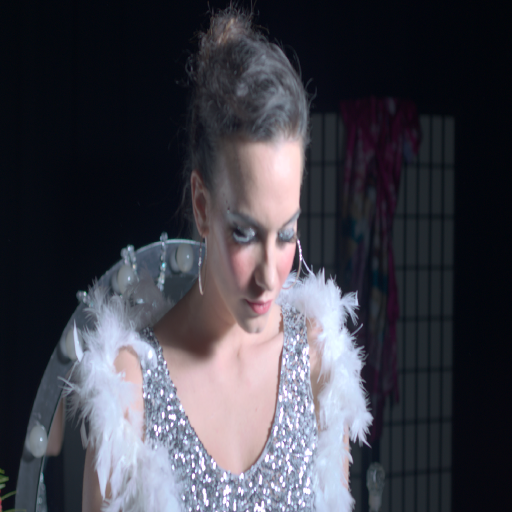} &
		\includegraphics[width=\widthface\textwidth,height=\widthfaceweight\textwidth]{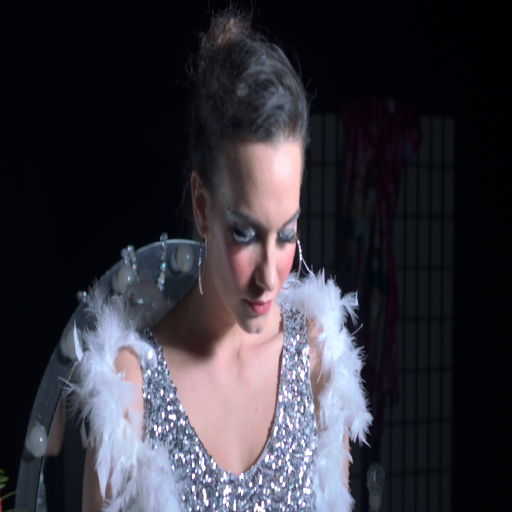} &
		\includegraphics[width=\widthface\textwidth,height=\widthfaceweight\textwidth]{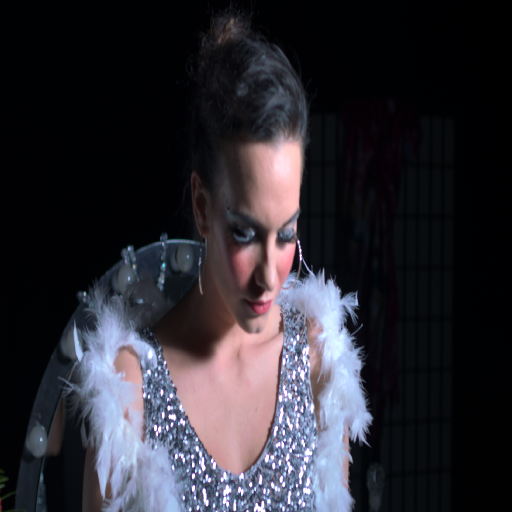} &
		\includegraphics[width=\widthface\textwidth,height=\widthfaceweight\textwidth]{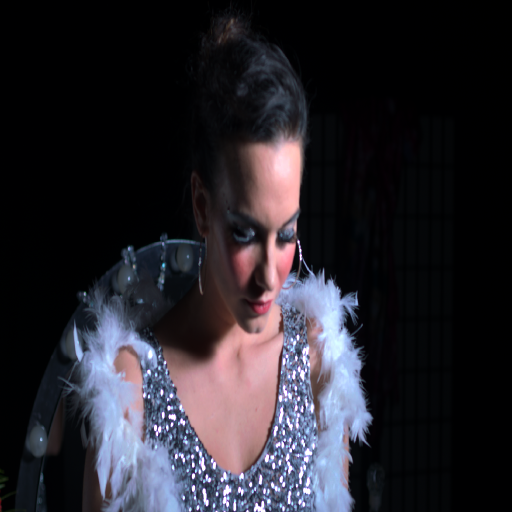} &
		\includegraphics[width=\widthface\textwidth,height=\widthfaceweight\textwidth]{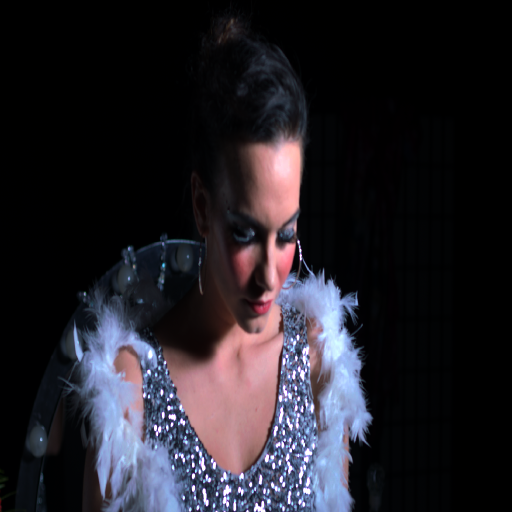} &
		\includegraphics[width=\widthface\textwidth,height=\widthfaceweight\textwidth]{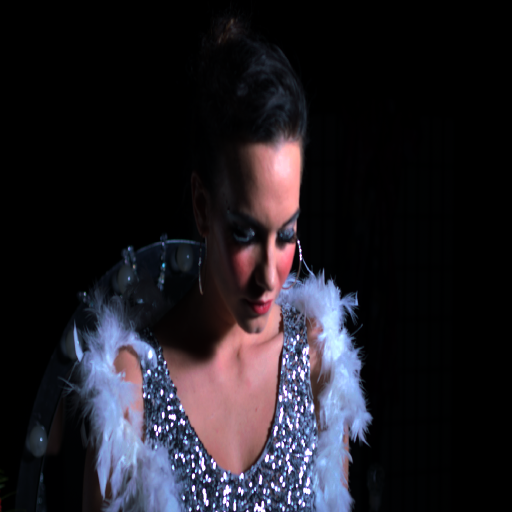} &
		\includegraphics[width=\widthface\textwidth,height=\widthfaceweight\textwidth]{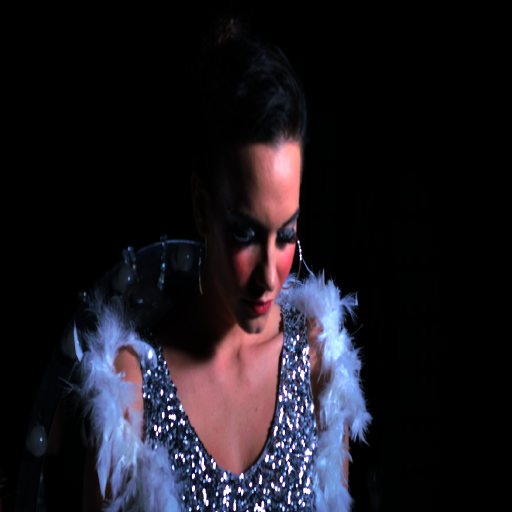} \\
		\multicolumn{8}{c}{LDR}\\
        &&&&&&&\\
		
		\includegraphics[width=\widthface\textwidth,height=\widthfaceweight\textwidth]{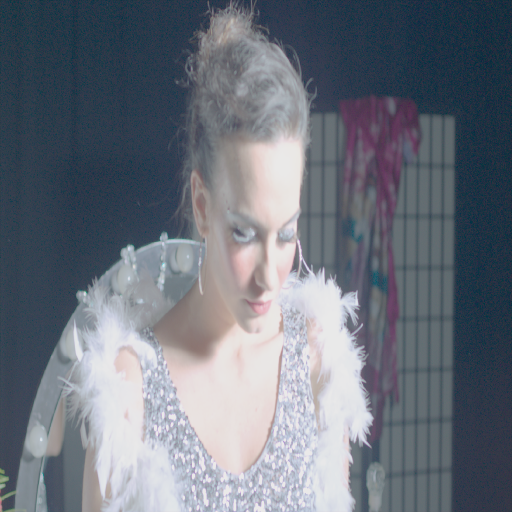} &
		\includegraphics[width=\widthface\textwidth,height=\widthfaceweight\textwidth]{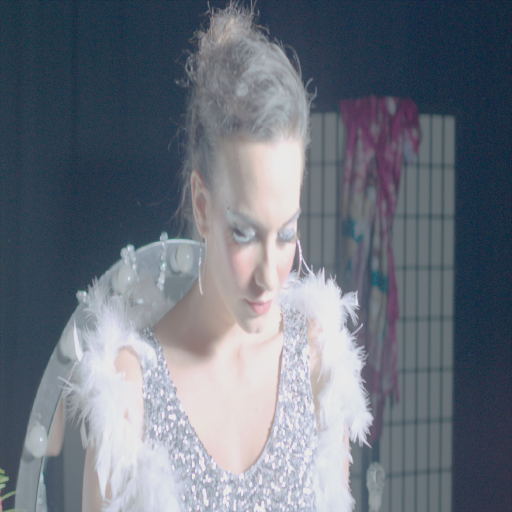} &
		\includegraphics[width=\widthface\textwidth,height=\widthfaceweight\textwidth]{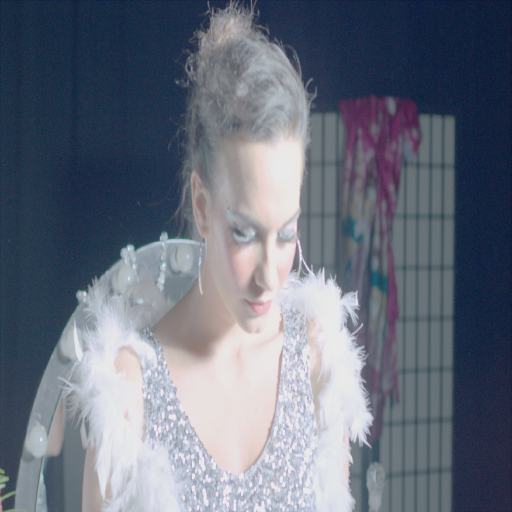} &
		\includegraphics[width=\widthface\textwidth,height=\widthfaceweight\textwidth]{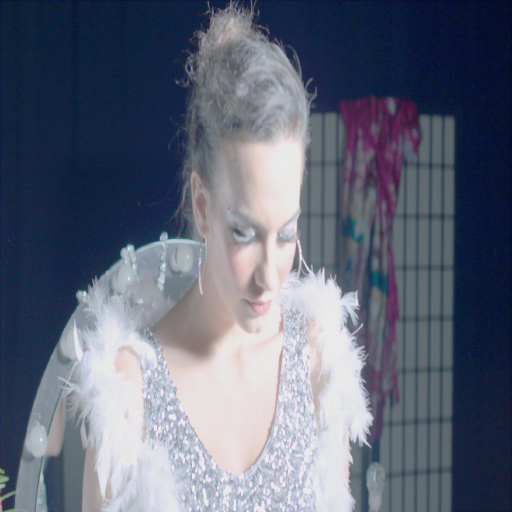} &
		\includegraphics[width=\widthface\textwidth,height=\widthfaceweight\textwidth]{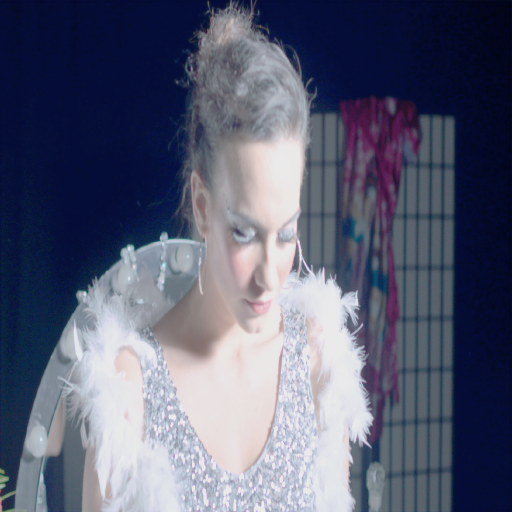} &
		\includegraphics[width=\widthface\textwidth,height=\widthfaceweight\textwidth]{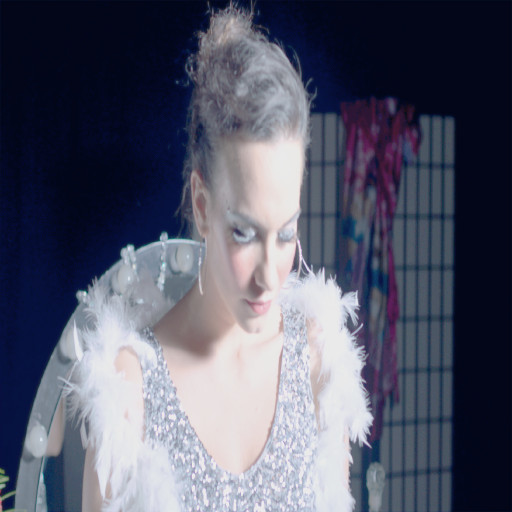} &
		\includegraphics[width=\widthface\textwidth,height=\widthfaceweight\textwidth]{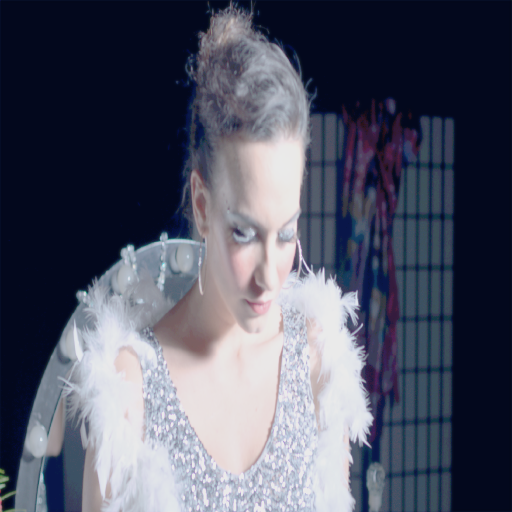} &
		\includegraphics[width=\widthface\textwidth,height=\widthfaceweight\textwidth]{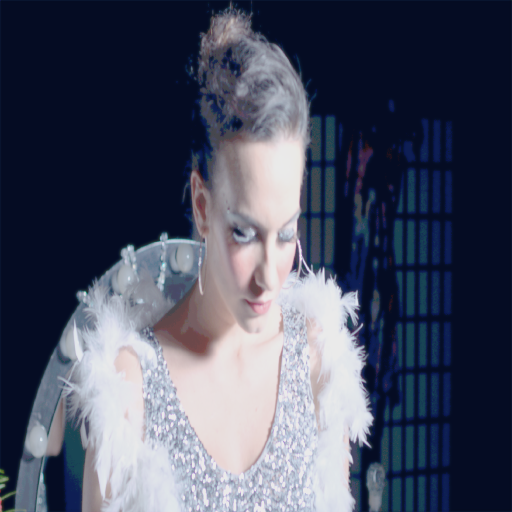} \\
		\multicolumn{8}{c}{ExpandNet \cite{marnerides2018expandnet}}\\
		&&&&&&&\\

		\includegraphics[width=\widthface\textwidth,height=\widthfaceweight\textwidth]{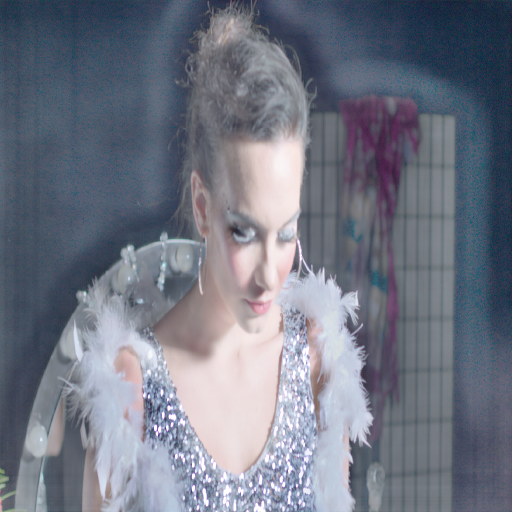}&
		\includegraphics[width=\widthface\textwidth,height=\widthfaceweight\textwidth]{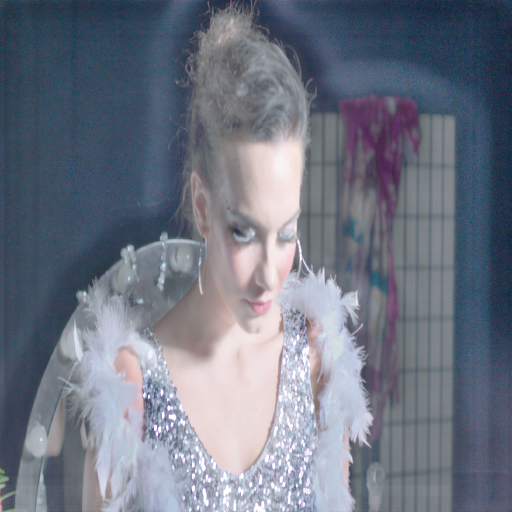} &
		\includegraphics[width=\widthface\textwidth,height=\widthfaceweight\textwidth]{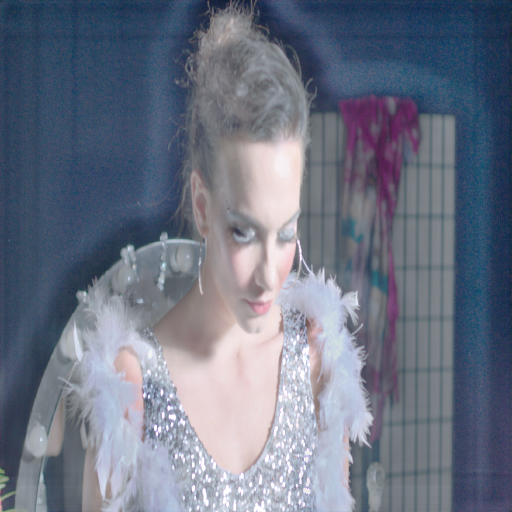} &
		\includegraphics[width=\widthface\textwidth,height=\widthfaceweight\textwidth]{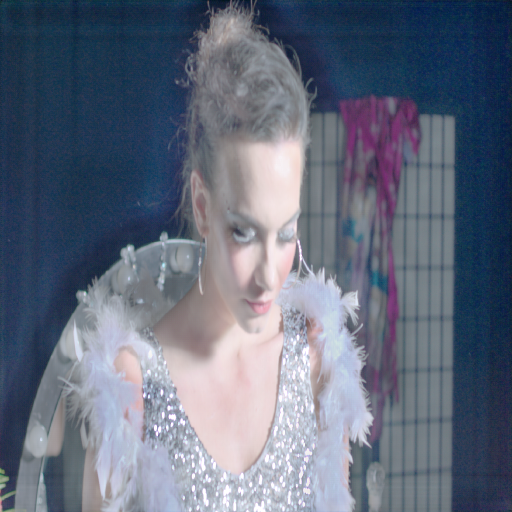} &
		\includegraphics[width=\widthface\textwidth,height=\widthfaceweight\textwidth]{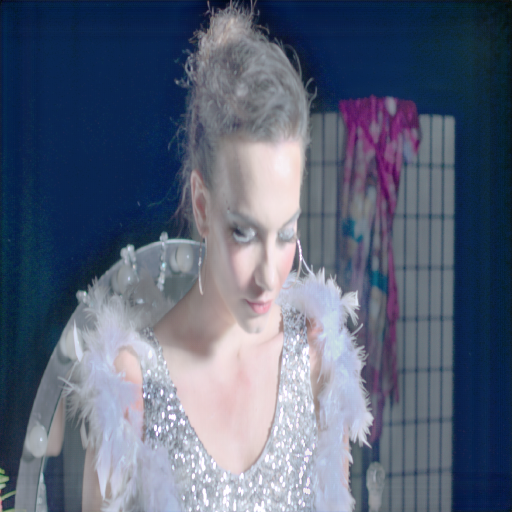} &
		\includegraphics[width=\widthface\textwidth,height=\widthfaceweight\textwidth]{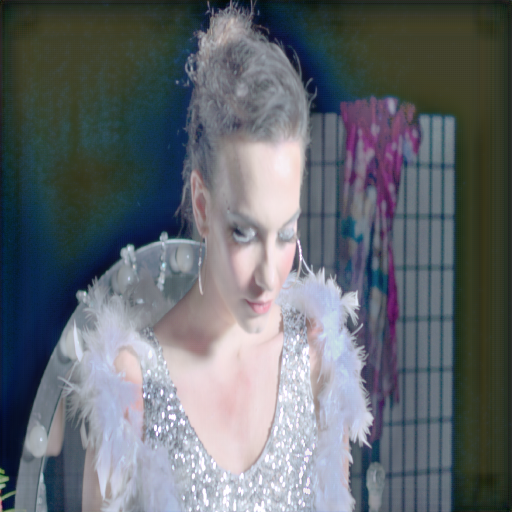} &
		\includegraphics[width=\widthface\textwidth,height=\widthfaceweight\textwidth]{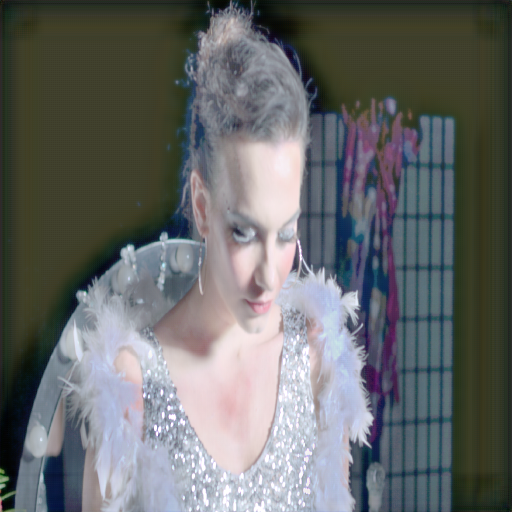} &
		\includegraphics[width=\widthface\textwidth,height=\widthfaceweight\textwidth]{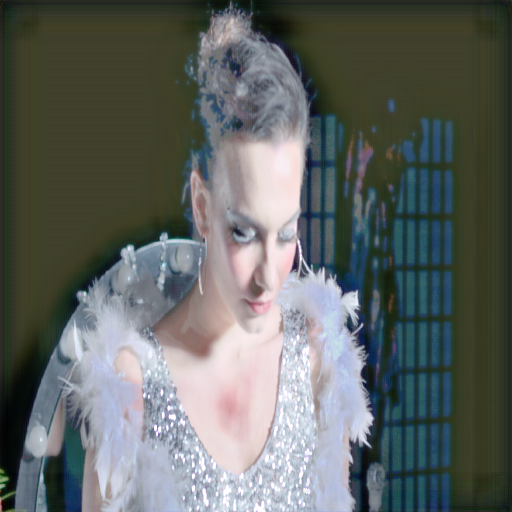} \\
		\multicolumn{8}{c}{HDRUNET \cite{chen2021hdrunet}}\\
		&&&&&&&\\
		
		\includegraphics[width=\widthface\textwidth,height=\widthfaceweight\textwidth]{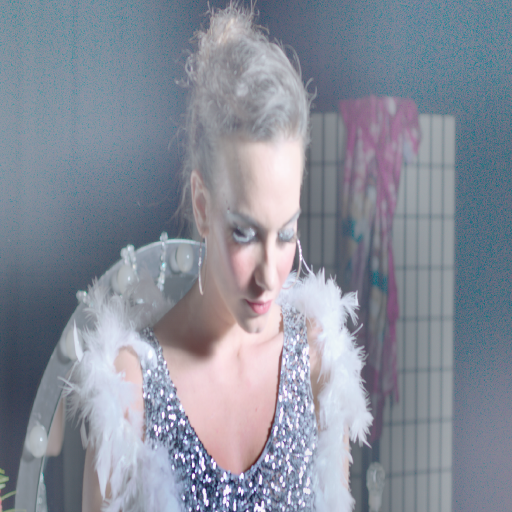} &
		\includegraphics[width=\widthface\textwidth,height=\widthfaceweight\textwidth]{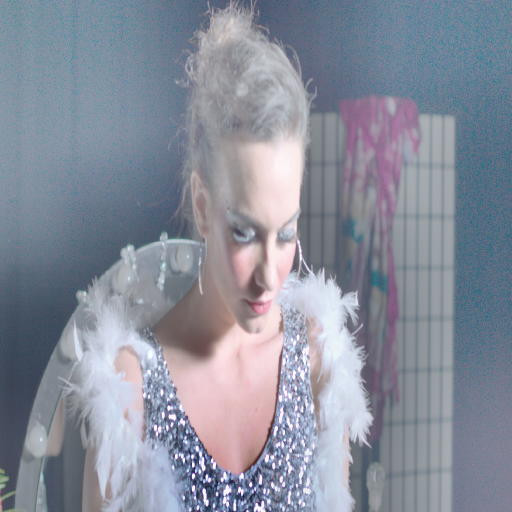} &
		\includegraphics[width=\widthface\textwidth,height=\widthfaceweight\textwidth]{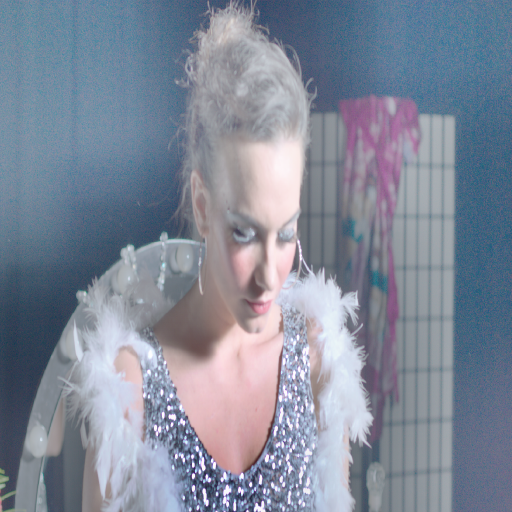} &
		\includegraphics[width=\widthface\textwidth,height=\widthfaceweight\textwidth]{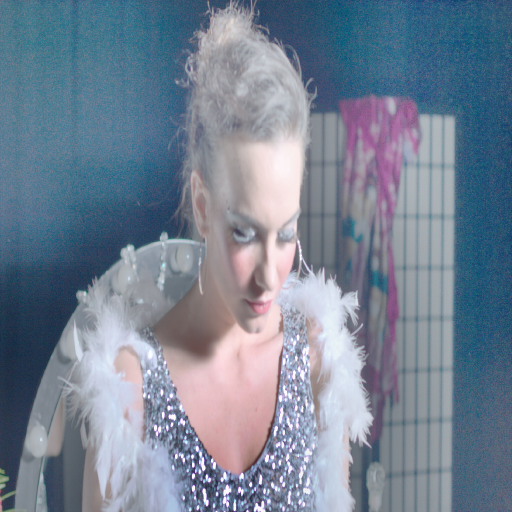} &
		\includegraphics[width=\widthface\textwidth,height=\widthfaceweight\textwidth]{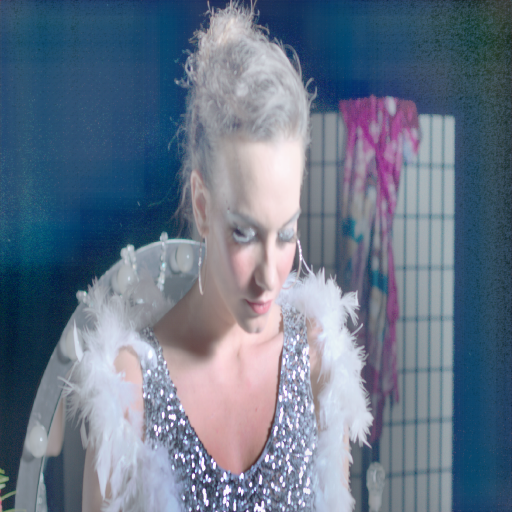} &
		\includegraphics[width=\widthface\textwidth,height=\widthfaceweight\textwidth]{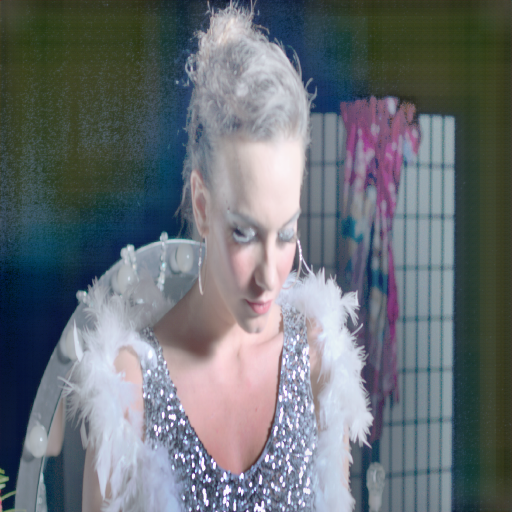} &
		\includegraphics[width=\widthface\textwidth,height=\widthfaceweight\textwidth]{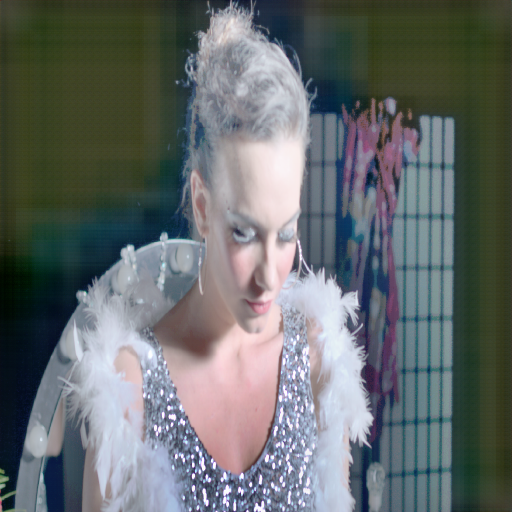} &
		\includegraphics[width=\widthface\textwidth,height=\widthfaceweight\textwidth]{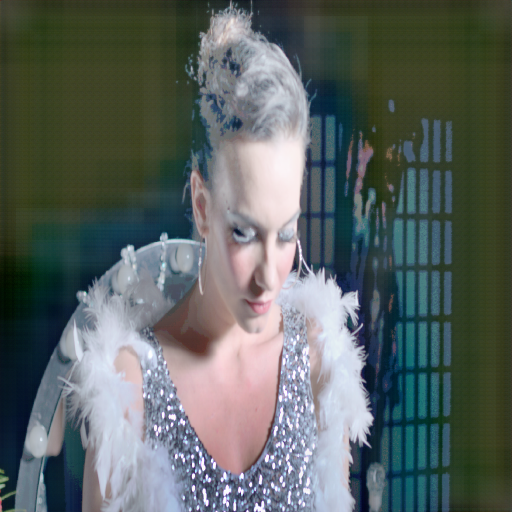} \\
		\multicolumn{8}{c}{Uformer \cite{Wang2021UformerAG}}\\
		&&&&&&&\\
		
		\includegraphics[width=\widthface\textwidth,height=\widthfaceweight\textwidth]{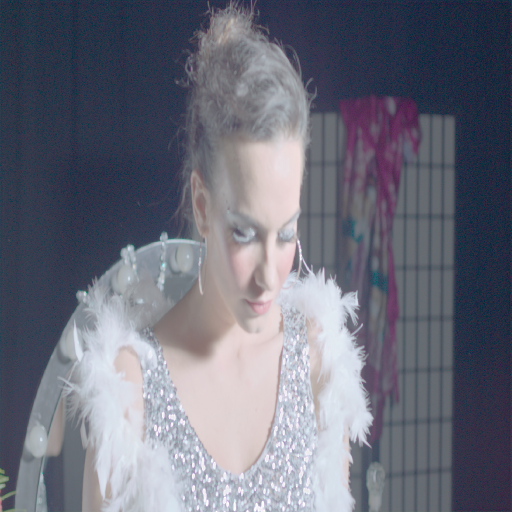} &
		\includegraphics[width=\widthface\textwidth,height=\widthfaceweight\textwidth]{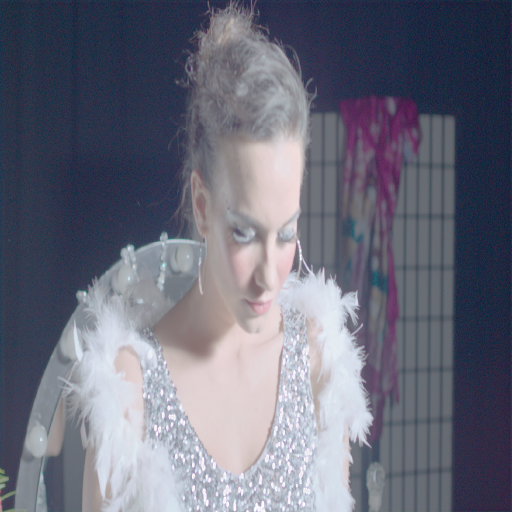} &
		\includegraphics[width=\widthface\textwidth,height=\widthfaceweight\textwidth]{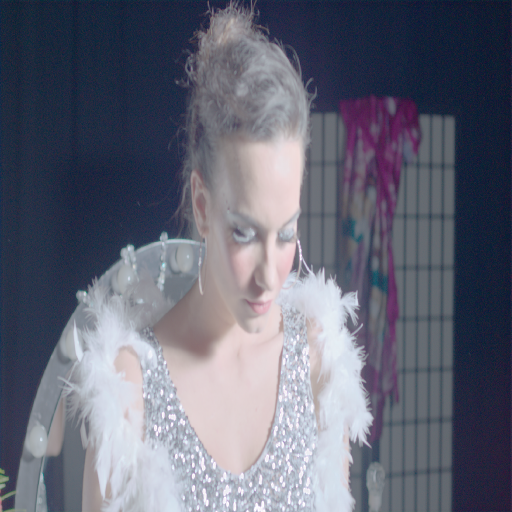} &
		\includegraphics[width=\widthface\textwidth,height=\widthfaceweight\textwidth]{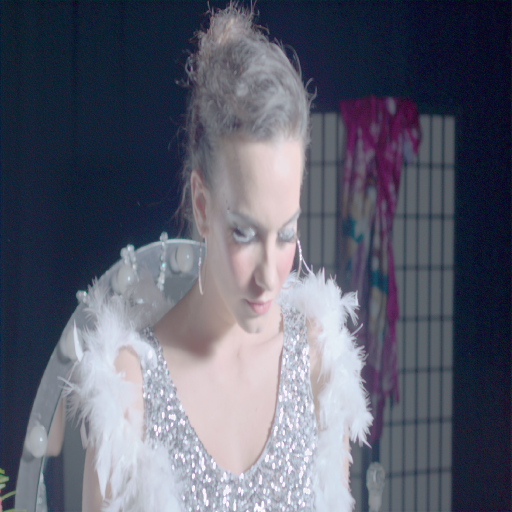} &
		\includegraphics[width=\widthface\textwidth,height=\widthfaceweight\textwidth]{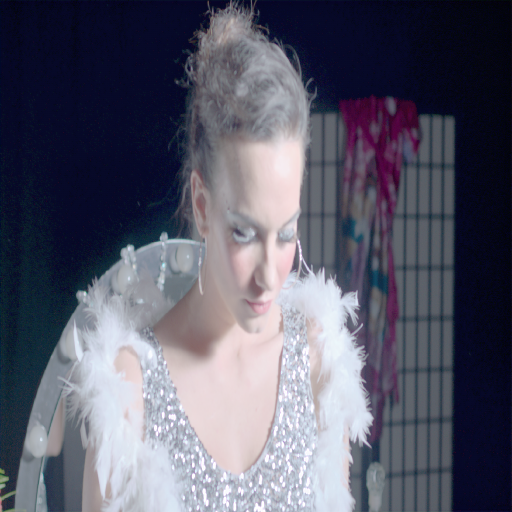} &
		\includegraphics[width=\widthface\textwidth,height=\widthfaceweight\textwidth]{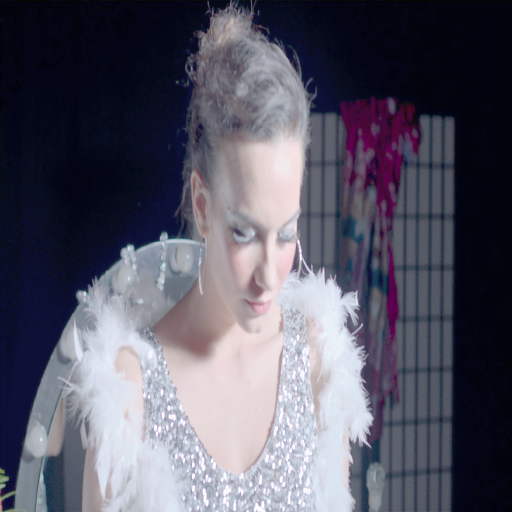} &
		\includegraphics[width=\widthface\textwidth,height=\widthfaceweight\textwidth]{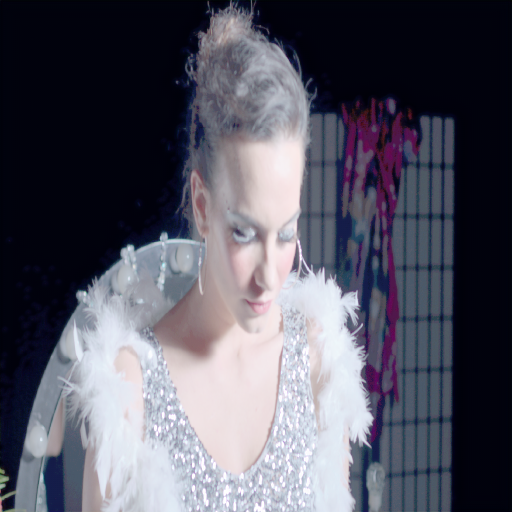} &
		\includegraphics[width=\widthface\textwidth,height=\widthfaceweight\textwidth]{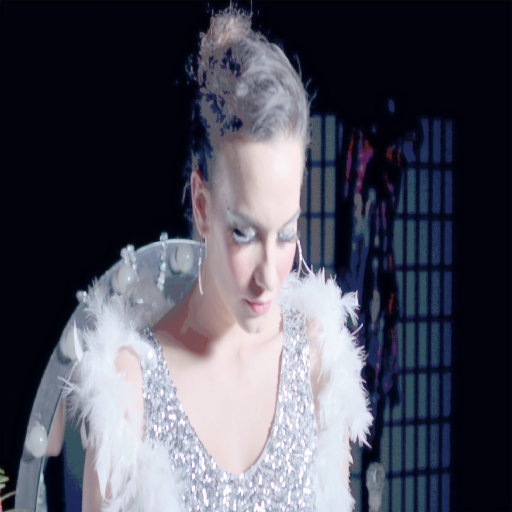} \\
		\multicolumn{8}{c}{Ours}\\
		&&&&&&&\\
		
		\includegraphics[width=\widthface\textwidth,height=\widthfaceweight\textwidth]{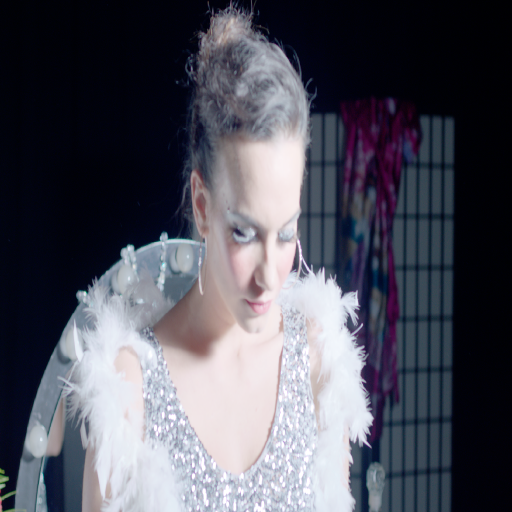} &
		\includegraphics[width=\widthface\textwidth,height=\widthfaceweight\textwidth]{Visualization_Sys/1321_tone_mapped_gt.png} &
		\includegraphics[width=\widthface\textwidth,height=\widthfaceweight\textwidth]{Visualization_Sys/1321_tone_mapped_gt.png} &
		\includegraphics[width=\widthface\textwidth,height=\widthfaceweight\textwidth]{Visualization_Sys/1321_tone_mapped_gt.png} &
		\includegraphics[width=\widthface\textwidth,height=\widthfaceweight\textwidth]{Visualization_Sys/1321_tone_mapped_gt.png} &
		\includegraphics[width=\widthface\textwidth,height=\widthfaceweight\textwidth]{Visualization_Sys/1321_tone_mapped_gt.png} &
		\includegraphics[width=\widthface\textwidth,height=\widthfaceweight\textwidth]{Visualization_Sys/1321_tone_mapped_gt.png} &
		\includegraphics[width=\widthface\textwidth,height=\widthfaceweight\textwidth]{Visualization_Sys/1321_tone_mapped_gt.png} \\
		\multicolumn{8}{c}{Ground Truth}\\
	\end{tabular}}
        \vspace{-0.2in}
	\caption{Visualization comparison in the Test-Gamma dataset, where HDR images are processed with gamma curves with various $\gamma$ values.}
        \vspace{-0.1in}
	\label{gamma}
\end{figure*}

\begin{figure*}[t]
	\centering
	\LARGE
	\newcommand\widthface{0.5}
	\resizebox{1.0\linewidth}{!}{
	\begin{tabular}{cccc}
	    $\gamma=0.2$ & $\gamma=0.3$  &$\gamma=0.4$ & $\gamma=0.5$\\
	    \includegraphics[width=\widthface\textwidth]{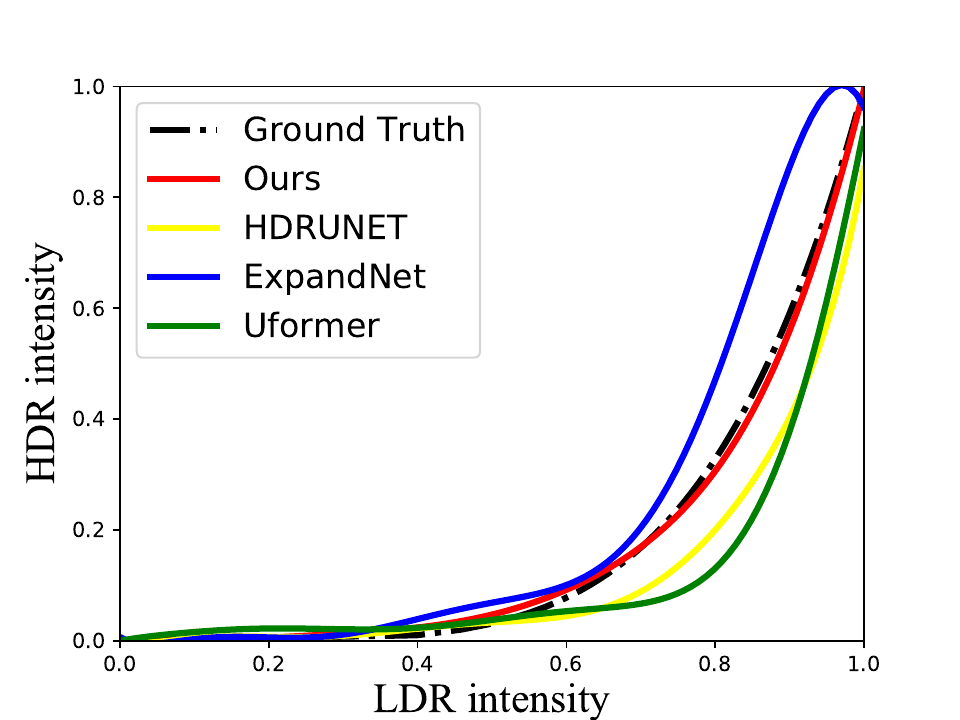} &
		\includegraphics[width=\widthface\textwidth]{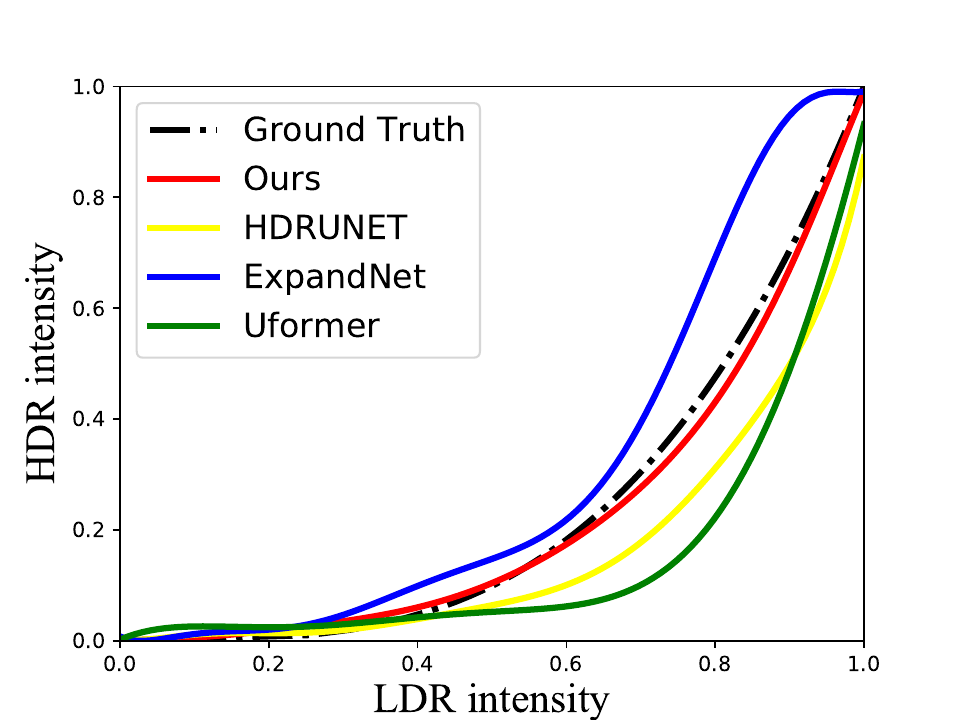} &
		\includegraphics[width=\widthface\textwidth]{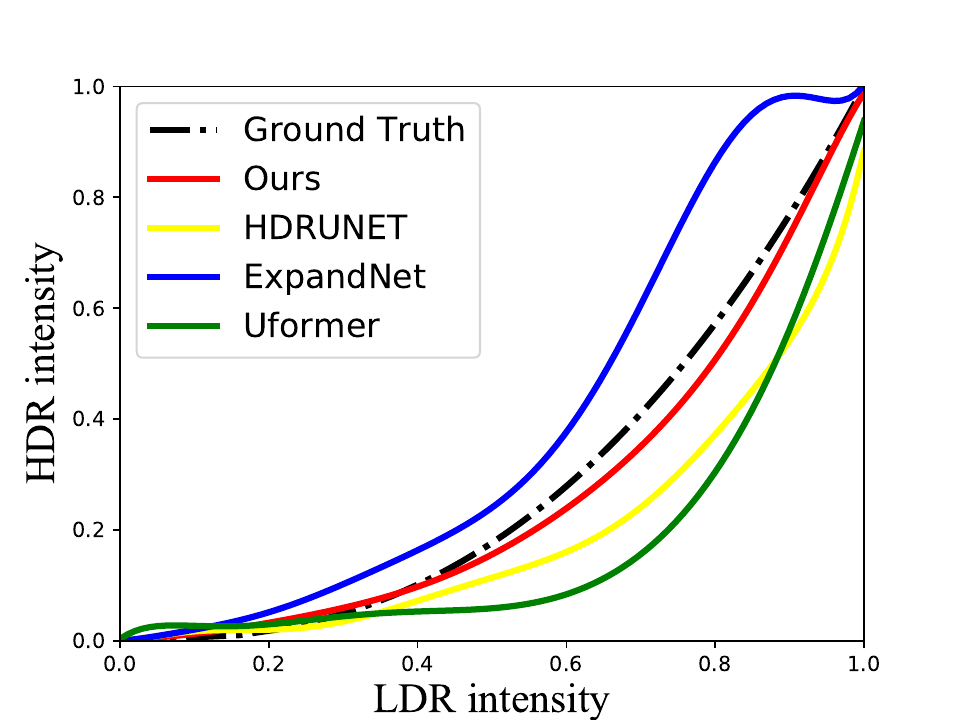} &
		\includegraphics[width=\widthface\textwidth]{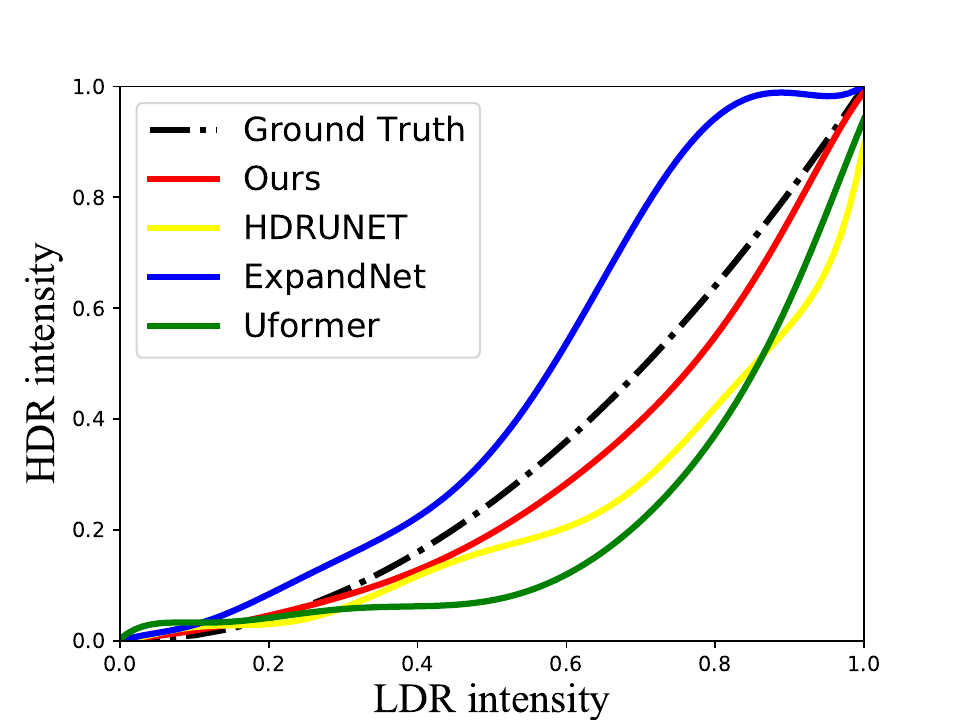} \\
		&&&\\
		
    	$\gamma=0.6$ & $\gamma=0.7$ & $\gamma=0.8$ &$\gamma=1.0$  \\
		\includegraphics[width=\widthface\textwidth]{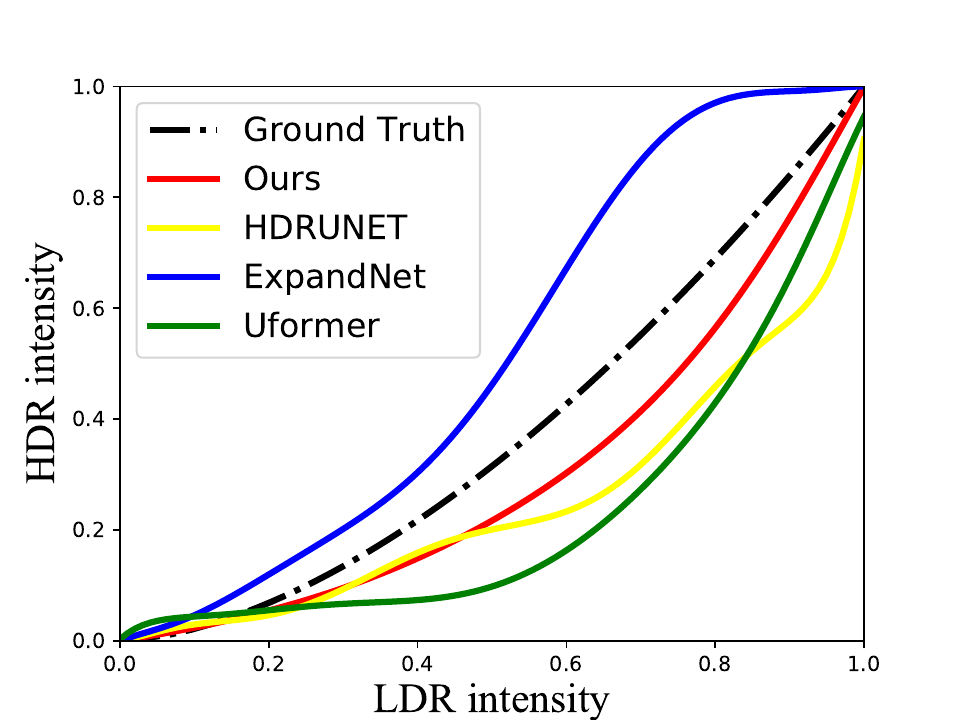} &
		\includegraphics[width=\widthface\textwidth]{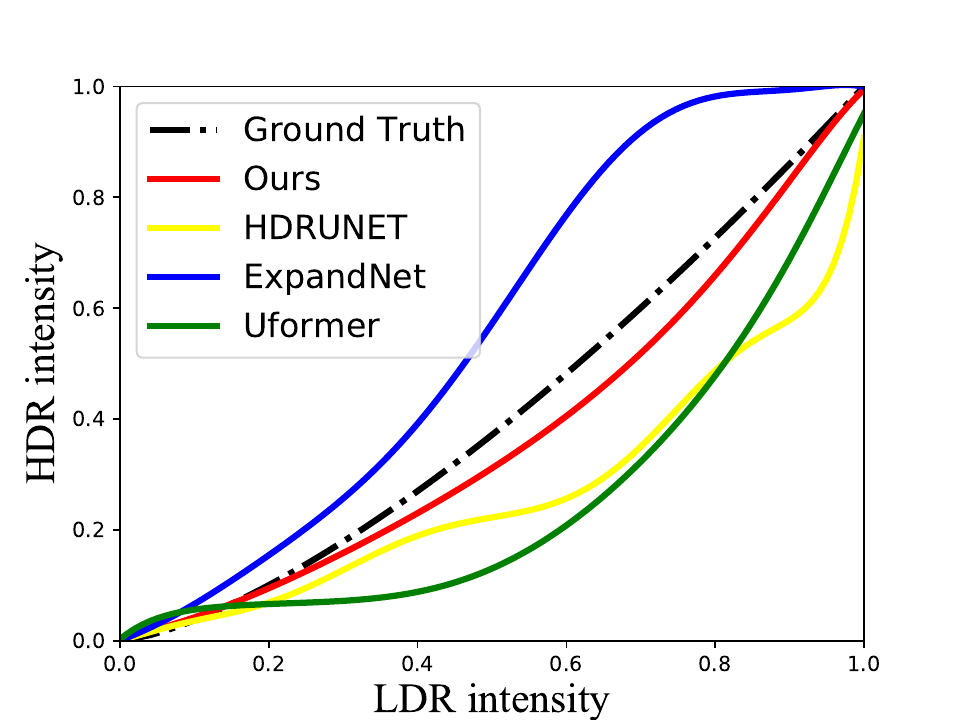} &
		\includegraphics[width=\widthface\textwidth]{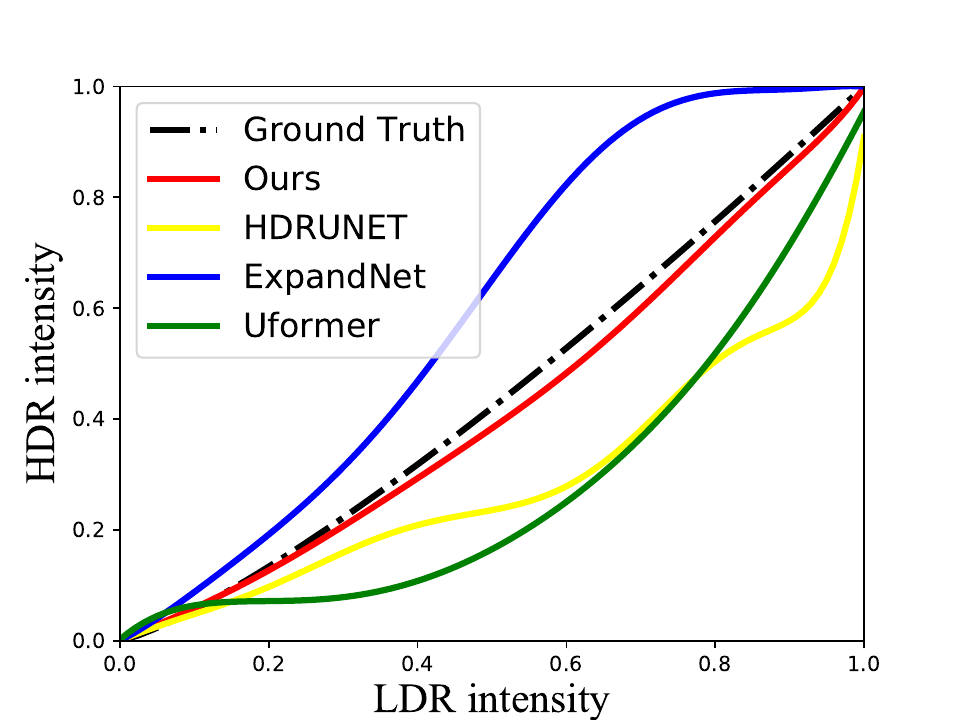} &
		\includegraphics[width=\widthface\textwidth]{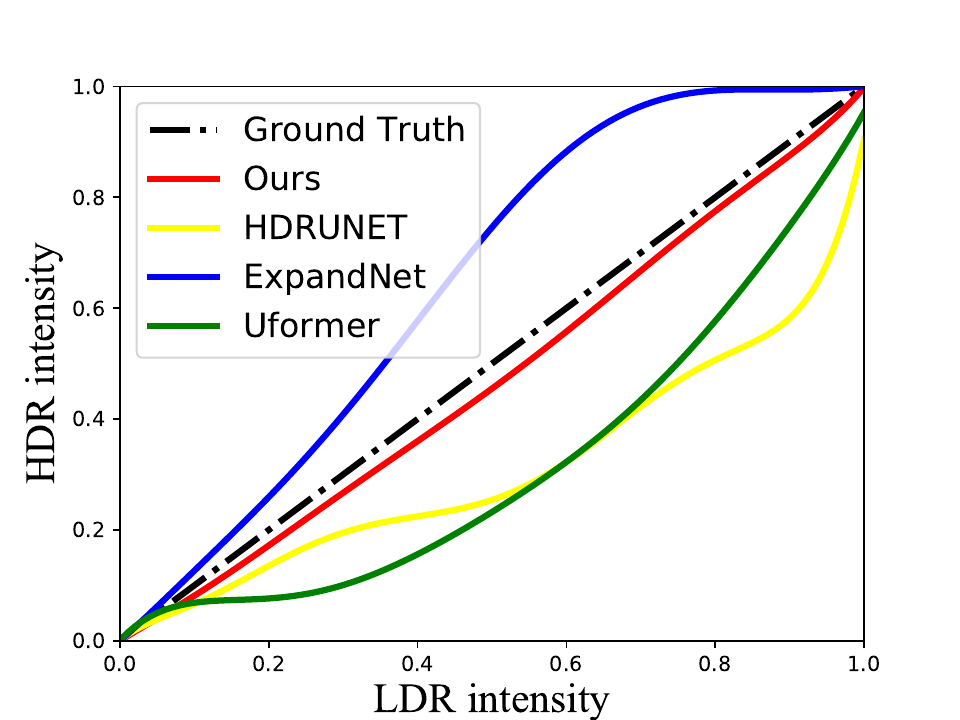} 
	\end{tabular}}
        \vspace{-0.2in}
	\caption{The evaluation of curve estimation in Test-Gamma dataset. Our method can explicitly recover the corresponding tone mapping function in a large tone space.}
 \vspace{-0.1in}
	\label{curve}
\end{figure*}

\section{Experiments}
\subsection{Quantitative Comparison} 
For the quantitative evaluation, we choose some SOTA methods for comparison.
The SOTA methods include two approaches that consider the implicit modeling of tone-mapping curves, i.e., ExpandNet~\cite{marnerides2018expandnet} and HDRUNET~\cite{chen2021hdrunet}. And recently, transformer-based frameworks have shown satisfactory performance in low-level vision tasks, e.g., image denoising and SR, and we also choose a general SOTA framework Uformer~\cite{Wang2021UformerAG} in the comparison.

We evaluate the performance of HDR reconstruction from two aspects, i.e., the HDR images themselves and the HDR images after tone-mapping functions. For the former, we compute the PSNR and SSIM of the reconstructed HDR images, and employ the metric of HDR-VDP-2.2 \cite{Narwaria2015HDRVDP22AC}. HDR-VDP-2.2 is specially designed for the HDR reconstruction task, and is tested with the following setting: HDR image is displayed on a 28-inch, 2K (2560*1440) screen, and the distance is 0.5 meters from the screen to the human eyes. For the evaluation after tone-mapping functions, we also compute the PSNR and SSIM values of images, so-called $\mu$-PSNR, and $\mu$-SSIM. 

Furthermore, to balance these two kinds of metrics, we design a new metric, AvgPSNR, which is defined as ${\rm AvgPSNR}=0.7\times {\rm PSNR}+0.3\times \mu{\rm -PSNR}$. AvgPSNR is a metric that can jointly assess the PSNR values of HDR images before and after tone mapping. We consider the evaluation before tone mapping since HDR reconstruction is our task's target, whose reconstruction accuracy is significant, and such a traditional PSNR value accounts for 70 percent of the AvgPSNR metric. On the other hand, we also consider the evaluation of HDR images in the human visual system. HDR images are usually displayed in 8-bit format after tone mapping on the common display devices for human perceptions. Thus, we introduce a 30 percent mu-PSNR into the metric to evaluate the visual effects of reconstructed HDR images on the standard displayed.
 
The results of the quantitative comparisons on the ``Test-Real'', which contain the real-world LDR-HDR pairs, are shown in Table~\ref{tab:qc}.
It can be seen that our method can achieve SOTA performance on all metrics except $\mu$-PSNR, showing the effectiveness of our approach to complete the HDR reconstruction task.
An anomaly occurs in ExpandNet \cite{marnerides2018expandnet}, which has a very low PSNR because ExpandNet cannot reconstruct HDR intensity images accurately.
But after tone-mapped, ExpandNet does a better job of showing the details in tone-mapped images.
Referring to the AvgPSNR metric, ExpandNet still cannot provide a good reconstruction capability for HDR reconstruction.
\subsection{Qualitative Comparison} 
First, we display the visual comparison on the ``Test-Real'' in Fig.~\ref{real}. It is clear that our method's results are closer to the ground truth, demonstrating the dynamic range expansion ability of our framework.
And we also provide the visualization to prove that our method can handle one scene with various tone-mapping curves. As shown in Fig.~\ref{gamma}, there are LDR images with gamma tone-mapping functions having different $\gamma$ values, and our method's results on all input LDR images approach the target ground truth. On the other hand, the results of other baselines are disparate from the ground truth, and the results are diverse for LDR images from the same scene while different tone-mapping curves. All the comparisons demonstrate the effectiveness of our framework in dealing with the HDR reconstruction task.

\begin{figure*}[t]
  \centering
  \includegraphics[width=1\textwidth]{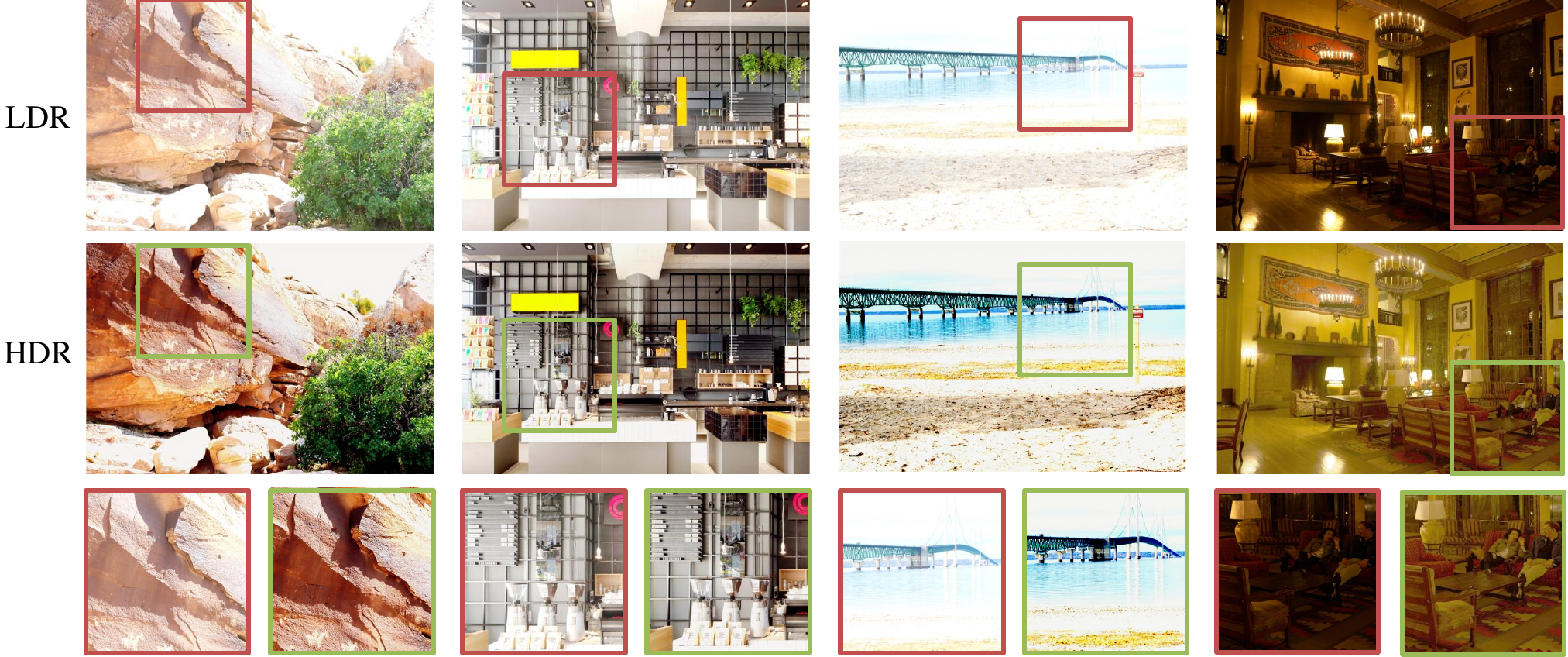}
  \vspace{-0.25in}
  \caption{The visual samples demonstrate that our framework can estimate the tone-mapping functions for LDR images in the wild. The bottom row contains the cropped areas for a clear comparison.}
  \label{fig:Wild}
  \vspace{-0.15in}
\end{figure*}

\subsection{Curve Estimation Evaluation}
One significant advantage of our framework is the explicit estimation of tone-mapping curves, and we conduct analysis on Test-Gamma to measure the curve estimation accuracy. Given an input LDR image, our framework estimates the polynomial coefficients for all pixels. To derive the global tone-mapping curve, we estimate the global polynomial coefficients by building the polynomial relation between the pixel values of the input LDR image and the generated HDR image with the least-squares method~\cite{lin2004radiometric}.
As for the approaches that employ the strategy of implicit curve modeling, we can also compute its implicitly learned curve via the least-squares method and the polynomial function assumption.
As shown in Fig.~\ref{curve}, the ground truths of the tone-mapping curves are gamma functions, while the estimated polynomial tone-mapping curves from our framework can perfectly be in accord with the ground truth curves. On the other hand, the curves of other baselines deviate from the ground truths. Thus, our framework can estimate the tone-mapping functions of LDR images, even though such curves have not appeared during the training.

\subsection{Ablation Study} \label{as}
In this section, we conduct ablation studies to verify the impact of significant components in our framework, i.e., PPViT's pyramid structure, IUTB, and LPE. 
To verify the effect of PPViT's pyramid structure, we perform the experiments where we delete all the pyramid paths in PPViT except the path with the input of $L$ shown in Fig.~\ref{fig:overall}. For the ablation study of IUTB, we replace the IUTB with the traditional transformer's forward module. And we also remove LPE from our framework to verify its effectiveness. 
The results are shown in Table~\ref{tab:ab} with names as ``w/o PPViT'', ``w/o LPE'', and ``w/o IUTB'', respectively, demonstrating that all these three components contribute to the HDR reconstruction performance of our framework.
Furthermore, we also set the experiments to prove the superiority of employing the smooth $L_1$ loss instead of the simple $L_1$ loss.
It is displayed in the column of ``$L_1$ loss'' in Table~\ref{tab:ab}, demonstrating that smooth $L_1$ loss can better optimize our model than $L_1$ loss.

\begin{table}[t]
  \centering
  \Large
  \caption{The results of ablation study in Test-Real dataset. \textcolor[rgb]{ 1,  0,  0}{Red} represents the setting with the best performance.}
  \resizebox{1.0\linewidth}{!}{
    \begin{tabular}{cccccc}
    \toprule
          & w/o PPViT & w/o LPE & w/o IUTB & $L_1$ loss & Ours \\
    \midrule
    PSNR  & 28.11 & 27.49 &  24.95 & 27.28 & \textcolor[rgb]{ 1,  0,  0}{28.24} \\
    SSIM  & 0.9365 & 0.9393 &   0.9121     & 0.9301 & \textcolor[rgb]{ 1,  0,  0}{{0.9406}} \\
    HDR-VDP-2.2 & 46.50      & 46.50 &   45.99    & 46.55 & \textcolor[rgb]{ 1,  0,  0}{{47.09}} \\
    AvgPSNR & 26.21 & 25.71 & 23.85     & 25.22 & \textcolor[rgb]{ 1,  0,  0}{{26.23}} \\
    \bottomrule
    \end{tabular}}%
  \label{tab:ab}%
\end{table}%
\subsection{HDR Reconstruction in the Wild}

As an approach that can effectively estimate the tone-mapping curves, our framework can be utilized to expand the dynamic range of LDR images in the wild. These images' tone-mapping curves could be largely different from the images in our experimental dataset. Especially, in order to further evaluate the effectiveness of our model in real scenes, we selected data from datasets with rich scenarios~\cite{cai2018learning}. The results are shown in Fig.~\ref{fig:Wild}, proving that our model can enhance details, color correction, and exposure reconstruction in the wild.

\section{Limitations}
Although our method can effectively estimate the tone mapping curve of a given LDR image, when extreme tone mapping curves (e.g., gamma=0.2) are encountered (although such a situation is a rare occurrence), our method cannot reconstruct a perfect HDR image. Our model will undergo inevitable color and light perturbations (more examples in supplementary materials). Adopting a more flexible tone curve estimation strategy and providing a larger tone space could be future work in HDR reconstruction.

\section{Conclusion}
In this paper, we explore the HDR reconstruction task where degradations are applied to HDR images with various tone-mapping functions. Through explicit modeling of tone-mapping curves, we propose a novel framework called Deep Polynomial Curve Estimation that can estimate the tone curves from the LDR images.
Extensive experiments on representative datasets are conducted, and our framework achieves SOTA performance, demonstrating the effectiveness of our approach.
As the first work for HDR reconstruction, our framework can be utilized for real-world HDR reconstruction in various scenarios.

In the future, there are two main directions to extend our work.
First, we plan to extend our framework for dealing with the MF-HDR reconstruction task, where the explicit curve estimation is also significant.
Different from the Si-HDR reconstruction task, the tone-mapping curves computed from multiple frames could be different, and the fusion mechanism is needed to merge the tone-mapping functions from varying frames. Moreover, as we have stated in this paper, the HDR reconstruction consists of two components, including recovering degenerated information and inverting the nonlinear mapping, and we have solved the latter part in this paper. In the future, it is worth exploring how to model the degradation in the pipeline from HDR to LDR images, e.g., dynamic range clipping, quantization, and noise, and then design the corresponding modules to invert the degradation.

\ack This work is partially supported by National Natural Science Foundation of China (No. 62206068). Besides, part of this work belongs to Jiaqi Tang's undergraduate thesis at Northwestern Polytechnical University.

\bibliography{egbib}
\end{document}